\documentclass[]{aastex631}

\usepackage{amsmath}
\usepackage[version=3]{mhchem}
\usepackage{appendix}

\usepackage{subfigure}

\received{March 29, 2025}
\revised{August 22, 2025}
\accepted{August 30, 2025}

\submitjournal{ApJ}

\begin{document}

\title{Planetary albedo is limited by the above-cloud atmosphere:\\Implications for sub-Neptune climate}

\author[0000-0002-2828-0396]{Sean Jordan}
\affiliation{ETH Zurich, Institute for Particle Physics \& Astrophysics, Wolfgang-Pauli-Str. 27, 8093 Zurich, Switzerland}
\correspondingauthor{Sean Jordan (jordans@ethz.ch)}

\author[0000-0002-8713-1446]{Oliver Shorttle}
\affiliation{Institute of Astronomy, University of Cambridge, UK}
\affiliation{Department of Earth Sciences, University of Cambridge, UK}

\author[0000-0003-3829-7412]{Sascha P. Quanz}
    \affiliation{ETH Zurich, Institute for Particle Physics \& Astrophysics, Wolfgang-Pauli-Str. 27, 8093 Zurich, Switzerland}
    \affiliation{National Center of Competence in Research PlanetS, Gesellschaftsstrasse 6, 3012 Bern, Switzerland}
    \affiliation{ETH Zurich, Department of Earth and Planetary Sciences, Sonneggstrasse 5, 8092 Zurich, Switzerland} 

\begin{abstract}

Energy limits that delineate the `habitable zone' for exoplanets depend on a given exoplanet's net planetary albedo (or `Bond albedo'). We here demonstrate that the planetary albedo of an observed exoplanet is limited by the above-cloud atmosphere\,---\,the region of the atmosphere that is probed in remote observation. We derive an analytic model to explore how the maximum planetary albedo depends on the above-cloud optical depth and scattering versus absorbing properties, even in the limit of a perfectly reflective grey cloud layer. We apply this framework to sub-Neptune K2-18b, for which a high planetary albedo has recently been invoked to argue for the possibility of maintaining a liquid water ocean surface, despite K2-18b receiving an energy flux from its host star that places it inside of its estimated `habitable zone' inner edge. We use a numerical multiple-scattering line-by-line radiative transfer model to retrieve the albedo of K2-18b based on the observational constraints from the above-cloud atmosphere. Our results demonstrate that K2-18b's observed transmission spectrum already restricts its possible planetary albedo to values below the threshold required to be potentially habitable, with the data favouring a median planetary albedo of 0.17\,--\,0.18. Our results thus reveal that currently characteriseable sub-Neptunes are likely to be magma-ocean or gas-dwarf worlds. The methods that we present are generally applicable to constrain the planetary albedo of any exoplanet with measurements of its observable atmosphere, enabling the quantification of potential exoplanet habitability with current observational capabilities.

\end{abstract}

\section{Introduction} \label{sec:intro}

Planetary albedo is a crucial parameter in determining the climates of habitable-zone exoplanets. Planetary albedo describes the fraction of incident starlight that is reflected away from an exoplanet's atmosphere via scattering. The remainder of the incident starlight not scattered away is instead absorbed and this absorbed energy determines the planet's equilibrium climate state. Absorbed energy fluxes (Wm$^{-2}$) delineate climatically catastrophic limits, such as the Runaway Greenhouse boundary. The mapping from energy-flux limits to orbital-space, or equilibrium temperature, limits of habitability therefore requires knowledge of the planetary albedo.

In principle, planetary albedo can vary between 0\,--\,1, however, in practice, not all planetary albedos are possible for a given planet--star system. Planetary albedo depends on the absorbing versus scattering properties of a planet's surface, atmospheric gases, and clouds/hazes, over the relevant wavelength range of the incident starlight. The range of wavelengths that are important for determining planetary albedo depends on the stellar type and effective temperature of the host star. Lower effective temperature stars, such as M-dwarfs, emit radiation at wavelengths further towards the near-infrared (NIR) compared to the emission of hotter stars like the Sun. Since Rayleigh scattering has a $\lambda^{-4}$ dependence, and molecular absorption bands of common atmospheric gases like \ce{CH4}, \ce{CO2}, and \ce{H2O}, occur at NIR and IR wavelengths, the starlight of M-dwarfs is less likely to be reflected and more likely to be absorbed in planetary atmospheres, compared to the starlight of G-dwarfs like the Sun. This is the reason why the habitable zone inner edge for Earth-like planets moves to lower instellation fluxes around lower effective temperature stars \citep{Kasting1993}.

Clouds and hazes tend to have stronger scattering than absorbing properties at wavelengths of peak stellar emission and can therefore increase the reflectivity of a planet compared to the reflectivity it would have from its surface and atmospheric gases alone. Water oceans, for example, have very low surface albedo, so overlying clouds cause a net increase in the planetary albedo. This creates the possibility that highly reflective cloud cover can mitigate the catastrophic climate transition of Runaway Greenhouse, up to high incident stellar fluxes. This line of reasoning has been recently used to argue that the sub-Neptune planet K2-18b, which orbits an M-dwarf host star and has been observed with JWST in transmission spectroscopy, could potentially host liquid water oceans \citep{MadhuCarbon}. This is despite K2-18b receiving an instellation flux of 1368\,Wm$^{-2}$, greatly in excess of the estimates for its Runaway Greenhouse threshold, estimated as 435\,Wm$^{-2}$ for the case of a 1\,bar \ce{H2}-dominated atmosphere, or 116\,Wm$^{-2}$ for the case of a 10\,bar \ce{H2}-dominated atmosphere \citep{Innes2023}. K2-18b is therefore an excellent case study for testing the role of albedo in establishing habitable conditions on exoplanets.

\subsection{K2-18b as a case study of sub-Neptune climate}

K2-18b is a 2.61$\pm0.09\,R_{\oplus}$ and 8.63$\pm1.35\,M_{\oplus}$ exoplanet \citep{Montet2015,Cloutier2017,Benneke2017} with a range of interior structures that would be consistent with its bulk density, from water-ocean world \citep{Madhu2020}, to magma-ocean world \citep{ShorttleMagma, FrancesMagma}, to mini-Neptune or `gas-dwarf' \citep{Wogan2024, greg}. K2-18b was one of 17 candidate planets from the K2 survey that was confirmed by \citet{Montet2015} and
highlighted for its intriguing bulk density, for receiving a similar incident stellar flux to that of the Earth, and for being optimally poised for atmospheric characterisation with transit spectroscopy thanks to orbiting a bright M-dwarf star. \citet{Benneke2019} obtained 9 transits of K2-18b with Hubble WFC3 and found that the transit spectrum showed evidence for \ce{H2O}, no evidence of \ce{CH4} or \ce{NH3} present at the pressure levels probed, \textcolor{black}{and constrained the cloud top pressure to 7.7\,--\,139\,mbar}. The \ce{H2O} detection and \ce{CH4} and \ce{NH3} non-detections were independently verified by other teams \citep{Madhu2020,Tsiaras2019}, \textcolor{black}{however \citet{Madhu2020} found no conclusive evidence for clouds or hazes, weakly constraining the possible cloud top to 0.1\,mbar\,--\,2\,bar.} \citet{Madhu2020} also proposed that the \ce{H2O} detection could indicate a habitable ocean world and that the role of potential alien biochemistry could not be ruled out for the \textcolor{black}{inferred} depletion of \ce{CH4} and \ce{NH3}. \textcolor{black}{\citet{Bezard2020}, in contrast, argued that \ce{CH4} absorption should dominate over \ce{H2O} absorption in the atmosphere of K2-18b, and the two molecules could be easily confused in the WFC3 data because of their spectral similarity between 1.1\,--\,1.55\,$\mu$m.}

Observations of K2-18b in transmission using JWST were obtained more recently by \citet{MadhuCarbon}. These subsequent observations revealed that the absorption feature previously interpreted as \ce{H2O} to 3.25\,$\sigma$ confidence \citep{Madhu2020}, should instead be attributed to \ce{CH4}\textcolor{black}{, as predicted by \citet{Bezard2020}}. \ce{CO2} was newly observed, \ce{H2O} now went undetected, and \ce{NH3} was once again not detected \citep{MadhuCarbon}. Based on the \ce{NH3} non-detection, and the results of previous studies into \ce{NH3} on sub-Neptunes \citep{Tsai2021,Hu2021,Yu2021}, \citet{MadhuCarbon} proposed that K2-18b was most likely to be a habitable ocean world with possible evidence of biological gases in the atmosphere. A reanalysis of the same JWST data by \citet{reanalysis}, using two different reduction pipelines and a wide range of data treatments, has since found evidence for \ce{CH4} at higher abundances than those found in \citet{MadhuCarbon} and no statistical evidence for \ce{CO2}. \textcolor{black}{\citet{MadhuCarbon} found only weak evidence for clouds or hazes, and constrained the possible cloud top pressure to lie below the observable photosphere ($\gtrsim$100\,mbar).} Clouds and hazes were also not found in the reanalysis \citep{reanalysis}. \textcolor{black}{These results thus raise} uncertainty over what K2-18b's planetary albedo may be.

The planetary albedo of K2-18b plays an important role in determining whether the liquid water ocean scenario is plausible. This is because the energy threshold for the onset of Runaway Greenhouse on sub-Neptunes, with low mean molecular weight atmospheres, is estimated to be far lower than the energy threshold for Runaway Greenhouse on Earth-like planets, with high mean molecular weight atmospheres. This difference in climate stability between Earth-like and sub-Neptune worlds is due to the effect of convective inhibition \citep{Innes2023,Leconte2024}. Convective inhibition operates by water vapour in a \ce{H2} background atmosphere generating super-radiative layers in which the temperature increases rapidly with pressure, causing the habitable zone inner edge to be farther away from a given host star as a function of the thickness of the \ce{H2} atmosphere. If K2-18b has an atmosphere of only 1\,bar of \ce{H2}, then \citet{Innes2023} report that the planet would need an albedo $\gtrsim$\,0.68 to maintain a surface water ocean, and if it has an atmosphere of 10\,bar of \ce{H2}, then K2-18b would require an albedo $\gtrsim$\,0.92 to maintain a surface water ocean. Convective inhibition by water vapour in a \ce{H2}-background atmosphere was later confirmed in a three-dimensionally resolved model by \citet{Leconte2024}, who calculated that K2-18b would need an albedo $\gtrsim$\,0.6 if it possessed a 1\,bar atmosphere, in order to maintain a surface water ocean. \citet{Leconte2024} also discuss how the high albedo requirement is inconsistent with the observed transmission spectrum: the dense haze required to achieve the requisite albedo would significantly flatten the spectrum compared to what is observed. \citet{MadhuCarbon} and \citet{greg} suggest that clouds underlying the observable region of K2-18b's atmosphere in transmission spectroscopy would be capable of generating a sufficiently high planetary albedo to deter the Runaway Greenhouse effect modelled by \citet{Innes2023} and \citet{Leconte2024}. In this paper, we show how the albedo of sub-Neptunes can be observationally estimated and apply our method to K2-18b\textcolor{black}{, using the retrieval constraints of \citet{MadhuCarbon}} to constrain its albedo and thus probable climate state.

\subsection{Constraining planetary albedo with transmission spectroscopy}

Given an observed transmission spectrum (as with K2-18b), we can place a limit on the maximum possible planetary albedo because absorption must be taking place above the cloud/haze layer in order to have generated features in the spectrum. Absorption is required to match any non-featureless transmission spectrum and constrains how high the albedo can be: incident stellar flux is deposited in the above-cloud atmosphere of a planet lowering the planetary albedo, even with a perfectly reflecting underlying cloud/haze layer. If we know the optical depth and single scattering albedo of the observed, above-cloud atmosphere of a planet, then we can calculate the range of planetary albedos that are possible. The stronger the atmospheric absorption of incident starlight is, the lower the maximum possible planetary albedo can be. 

Previous investigation into the absorption of incident stellar radiation in the upper atmospheres of exoplanets has generally focused on planets with inverted temperature profiles \citep[e.g.,][]{Guillot2010}. Atmospheric temperature inversions occur when the heating rates due to the stellar radiation are greater than the heating rates due to the planet's own emission, for example due to the presence of TiO and VO at altitude in an exoplanet's atmosphere \citep{Burrows2007,Guillot2010}. Standard greenhouse gases in planetary atmospheres, however, such as \ce{CH4}, \ce{CO2}, and \ce{H2O}, absorb some stellar radiation but still \textcolor{black}{absorb more strongly at the wavelengths of the planetary emission}. Molecular absorption of starlight by common atmospheric greenhouse gases therefore does not necessarily lead to inverted temperature profiles, but does limit the range of planetary albedos that a cloudy planet can possibly attain. The limit on the attainable albedo depends on the degree of absorption of stellar radiation above the clouds.

\begin{figure}[b!]
    \centering
    \includegraphics[width=\columnwidth]{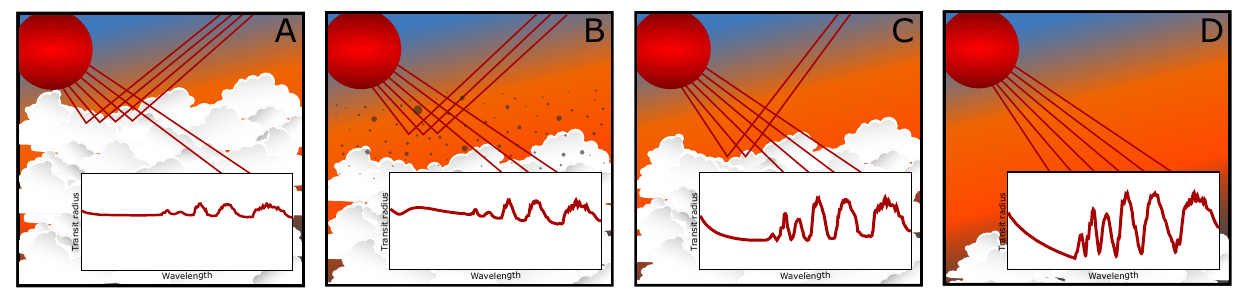}
    \caption{An observed transmission spectrum is linked to the planetary albedo through the optical depth, $\tau_{\infty}$, from the cloud top to the top of the atmosphere, and the scattering albedo, $w_0 = \tau_{\rm sca}/(\tau_{\rm abs}+\tau_{\rm sca})$, of the above-cloud atmosphere. High-altitude clouds/hazes that truncate observations and lead to flat transmission spectra can achieve high planetary albedos because there is little opportunity for stellar energy to be deposited above the clouds/haze (low $\tau_{\infty}$, panel A). If clouds/hazes exist deeper in planetary atmospheres, the planetary albedo depends on the scattering and absorption properties of the above-cloud atmosphere: if the atmosphere is very highly scattering then the albedo will be high and the transmission spectrum will be truncated due to the scattering continuum (high $w_0$, panel B); if the atmosphere is not highly scattering then some amount of the incident stellar energy must be deposited in the above-cloud atmosphere resulting in spectral absorption features, cloud truncation at more transparent wavelengths, and a relatively low planetary albedo (low $w_0$, panel C). If clouds or hazes are not visible in the transmission spectrum then the albedo will be low, due to the combination of surface reflection and Rayleigh scattering of the atmospheric gases (high $\tau_{\infty}$, panel D).}
    \label{fig:schematic}
\end{figure}

The absorption of stellar radiation above the clouds, and thus the calculable limit on albedo, depends on the optical depth and scattering/absorbing properties of the above-cloud atmosphere (Figure \ref{fig:schematic}). If high altitude clouds are present, the planet will have a low above-cloud optical depth, resulting in a high planetary albedo (approaching the albedo of the cloud material itself) and shallow spectral absorption features against the cloud continuum (Figure \ref{fig:schematic}A). If clouds exist at deeper pressures, the planet will have a higher above-cloud optical depth: in this case, the albedo and spectral appearance depends on the ratio of scattered to absorbed light (defined as $w_0 = \tau_{\rm sca}/(\tau_{\rm abs}+\tau_{\rm sca})$). If scattering dominates strongly over absorption due to extensive above-cloud hazes, then, even with a deep cloud layer, the planet can potentially still have a high albedo. However, in this case the transit spectrum will show evidence of scattering features with heavily muted absorption features (Figure \ref{fig:schematic}B). If absorption dominates, then the planet will have a lower albedo and the transit spectrum will show strong absorption features with cloud truncation in the intervening troughs (Figure \ref{fig:schematic}C). If clouds and hazes are not substantial at the pressure levels probed, then the albedo will be low (approaching that due to the combination of surface reflection and atmospheric Rayleigh scattering) and the transit spectrum will show deep and non-truncated absorption features (Figure \ref{fig:schematic}D).

\subsection{Paper structure}

In this paper, we formalise the link between the observed `above-cloud' atmosphere of an exoplanet and its maximum attainable planetary albedo. We derive an analytical formula to demonstrate how reflection from a perfectly reflective cloud layer produces net planetary albedos that still depend on the absorption and scattering properties of the above-cloud atmosphere, and demonstrate the behavior numerically with a multiple-scattering radiative-transfer routine. We apply our method to the much-debated example of K2-18b as a case study, and demonstrate how the retrieval constraints on K2-18b's observable atmosphere \textcolor{black}{from \citet{MadhuCarbon}} already limit its planetary albedo to values lower than the threshold for maintaining a liquid water ocean \textcolor{black}{\citep{Innes2023,Leconte2024}}. We end with discussion on K2-18b as a likely magma ocean or gas-dwarf world, and the broader application of the methods presented in this work.

\section{Methods \label{sec:methods}}

We first derive the analytic expression for maximum planetary albedo under the two-stream approximation. We later use a fully multiple-scattering line-by-line numerical radiative transfer method.

\subsection{Analytic derivation of maximum planetary albedo \label{sec:methods_analytic}}

We begin with the two stream equations \citep{TheBible}:

\begin{align}
    \frac{d}{d\tau}I_+ \, &= \, - \, \gamma_1 \, I_+ \,+\, \gamma_2 I_- + \gamma_B \, \pi \, B \, + \, \gamma_+ \, L \, \exp \left(-\frac{(\tau_\infty \, - \, \tau)}{\cos\zeta}\right),
    \\
    \frac{d}{d\tau}I_- &= \, \gamma_1 \, I_- \,-\, \gamma_2 I_+ - \gamma_B \, \pi \, B \, - \, \gamma_- \, L \, \exp \left(-\frac{(\tau_\infty \, - \, \tau)}{\cos\zeta}\right),
\end{align}

where

\begin{align}
    \gamma_1 \, &= \, \gamma \, \left(1 \, - \, g \, w_0 \right) + \gamma' \, \left(1 \, - \, w_0 \right),
    \\
    \gamma_2 \, &= \, \gamma \, \left(1 \, - \, g \, w_0 \right) - \gamma' \, \left(1 \, - \, w_0 \right),
    \\
    \gamma_B \, &= \, 2 \, \gamma' \, \left(1 \, - \, w_0 \right),
    \\
    \gamma_+ \, &= \, \frac{1}{2} \, w_0 \, - \, \gamma \, w_0 \, g \, \cos\zeta,
    \\
    \gamma_- \, &= \, \frac{1}{2} \, w_0 \, + \, \gamma \, w_0 \, g \, \cos\zeta.
\end{align}

for \textcolor{black}{upward radiation flux, $I_{+}$, downward radiation flux, $I_{-}$,} zenith angle, $\zeta$, vertical optical depth (increasing upwards), $\tau$, asymmetry factor, $g$, and closure coefficients, $\gamma$ and $\gamma'$ \citep{TheBible}.

Equations (1) and (2) are combined to to obtain the approximate formulas for the net vertical flux (equation 8) and actinic flux (equation 9) respectively:

\begin{align}
    \frac{d}{d\tau} \, \left( I_+ \, - \, I_- \right) \, &= \, - \, \left( \gamma_1 \, - \, \gamma_2 \right) \, \left( I_+ \,+\, I_- \right) \, + \, 2 \, \gamma_B \, \pi \, B \, + \, \left( \gamma_+ + \gamma_- \right) \, L \, \exp \left(-\frac{(\tau_\infty \, - \, \tau)}{\cos\zeta}\right),
    \\
    \frac{d}{d\tau} \, \left( I_+ \, + \, I_- \right) \, &= \, - \, \left( \gamma_1 \, + \, \gamma_2 \right) \, \left( I_+ \,-\, I_- \right) \, + \, \left( \gamma_+ - \gamma_- \right) \, L \, \exp \left(-\frac{(\tau_\infty \, - \, \tau)}{\cos\zeta}\right).
\end{align}

We can solve the two stream equations to obtain an expression for the planetary albedo as a function of the total optical depth of the atmosphere, accounting for both scattering and absorption, controlled by the parameter $w_0$. In this example, we are interested in the maximal possible reflection of incident starlight back out of the top of the atmosphere from a cloud layer resting at some finite optical depth  from the top of the atmosphere. This can be calculated directly based on retrieved physical and chemical constraints of the observable atmospheres of exoplanets. We will therefore ignore the planet's own emission (the term in $B$), and take $g = 0$ and $w_0 = const.$ to describe the above-cloud atmosphere. As in \citet{TheBible} one proceeds by first taking the derivative of equation (9) and substituting $\frac{d}{d\tau}\left(I_+-I_-\right)$ from equation (8):

\begin{align}
    \frac{d^2}{d\tau^2} \, \left( I_+ \, + \, I_- \right) \, &= \, - \, \left( \gamma_1 \, + \, \gamma_2 \right) \, \frac{d}{d\tau}\left( I_+ \,-\, I_- \right) \, + \, \frac{\left( \gamma_+ - \gamma_- \right)}{\cos\zeta} \, L \, \exp \left(-\frac{(\tau_\infty \, - \, \tau)}{\cos\zeta}\right),
    \\
    \frac{d^2}{d\tau^2} \, \left( I_+ \, + \, I_- \right) \, &= \, K^2 \, \left( I_+ \,+\, I_- \right) \, - \, 2\, \gamma \, w_0 \, L \, \exp \left(-\frac{(\tau_\infty \, - \, \tau)}{\cos\zeta}\right),
\end{align}

where ${K \, = \, 2 \, \sqrt{\gamma \, \gamma' \, \left( 1 \, - \, g \, w_0 \right) \, \left( 1 \, - \, w_0 \right)} } = \, 2 \, \sqrt{\gamma \, \gamma' \, \left( 1 \, - \, w_0 \right) }$ with ${ \rm g \, = \, 0 }$.

The solution to this differential equation requires summing the particular solution and superpositions of the two homogeneous solutions \citep{TheBible}:
\begin{align}
    I_+ \, + \, I_- \, = \, - \, \frac{2 \, \gamma \, w_0 \, L \, \cos^2\zeta}{1 \, - \, K^2 \, \cos^2\zeta} \, \exp \left(-\frac{(\tau_\infty \, - \, \tau)}{\cos\zeta}\right) \,
    + \, a \, \exp \left( K\left( \tau_\infty \, - \, \tau \right) \right)
    + \, b \, \exp \left( - \, K\left( \tau_\infty \, - \, \tau \right)\right)
\end{align}
and solving for expressions for $a$ and $b$ using two boundary conditions.

First, one must obtain two expressions in terms of \textcolor{black}{${I_+, I_-, a, \, {\rm and} \, b}$}. Equation (12) provides the first; the second is obtained by differentiating equation (12) and eliminating $\frac{d}{d\tau}\left(I_++I_-\right)$ from equation (9):

\begin{align}
    - \, 2 \, \gamma \, \left( I_+ \, - \, I_- \right) \, = \, - \, \frac{2 \, \gamma \, w_0 \, L \, \cos\zeta}{1 \, - \, K^2 \, \cos^2\zeta} \, \exp \left(-\frac{(\tau_\infty \, - \, \tau)}{\cos\zeta}\right) \,
    - \, a \, K \, \exp \left( K\left( \tau_\infty \, - \, \tau \right) \right)
    + \, b \, K \, \exp \left( - \, K\left( \tau_\infty \, - \, \tau \right)\right).
\end{align}

Using the boundary condition that the downward diffuse flux must be zero at the top of the atmosphere (${ \rm \tau = \tau_\infty }$):

\begin{align}
    I_-(\tau_\infty) \, &= \, 0,
    \\
    a \, \left( 2 \, \gamma \, - \, K \right) \, + \, b \, \left( 2 \, \gamma \, + \, K \right) \, &= \, \frac{2 \, \gamma \, w_0 \, L \, \cos\zeta \, \left( 2 \, \gamma \, \cos\zeta \, + \, 1 \right)}{1 \, - \, K^2 \, \cos^2\zeta}.
\end{align}

The second boundary condition is to impose upward reflection at the cloud top (i.e., the base, ${ \rm \tau = 0 }$, of the above-cloud atmosphere):
\begin{align}
    I_+(0) \, = \, \alpha_0 \, \left( I_-(0) \, + \, L \, \cos\zeta \, \exp \left(-\frac{(\tau_\infty)}{\cos\zeta}\right) \right),
\end{align}
\textcolor{black}{for albedo, $\alpha_0$}. For total perfect reflection at the cloud top, this becomes:

\begin{align}
    I_+(0) \, = \, I_-(0) \, + \, L \, \cos\zeta \, \exp \left(-\frac{(\tau_\infty)}{\cos\zeta}\right), \label{eq:BC_perfect_reflection}
\end{align}

By imposing perfect reflection at the cloud top, we obtain a fundamental upper limit on the maximum planetary albedo as a function of ${\rm \tau_\infty}$ and ${w_0}$ in the above-cloud atmosphere only. The parameters from the below-cloud atmosphere need not be considered because they can now only ever reduce the planetary albedo compared to the upper limit imposed by the above-cloud atmosphere, which can be constrained observationally. The boundary condition immediately eliminates ${\rm I_+}$ and ${\rm I_-}$ from equation (13):

 \begin{align}
     a \, K \, \exp(K\,\tau_\infty) - b \, K \, \exp(-\,K\,\tau_\infty) = 2 \, \gamma \, L \, \cos\zeta \, \left(1 \, - \, \frac{w_0}{1 \, - \, K^2 \, \cos^2\zeta}\right) \, \exp\left( -\frac{\tau_\infty}{\cos\zeta} \right).
 \end{align}

Equations (15) and (18) are now simultaneous equations in $a$ and $b$ from which it is trivial, but somewhat algebraically tedious, to obtain expressions for each of $a$ and $b$ respectively. The expressions for $a$ and $b$ thus obtained are:

 \begin{align}
     a \, &= \, L \, \cos\zeta \, \left( \frac{2 \, \gamma \, w_0 \, (2 \, \gamma \, \cos\zeta \, + \, 1)}{(2 \, \gamma \, - \, K)\,(1 \, - \, K^2 \, \cos^2\zeta)} \, - \, \frac{2 \, \gamma \, (2 \, \gamma \, + \, K)}{(2 \, \gamma \, - \, K)} \, C \right),
     \\
     b \, &= \, L \, \cos\zeta \, \left( 2 \, \gamma \, C \right),
 \end{align}

where we have defined the expression $C$, for brevity, as:

 \begin{align}
     C \, = \, \left( \frac{ \frac{w_0(2\,\gamma\,\cos\zeta \, + \, 1)}{(2\,\gamma \, - \, K)\,(1 \, - \,K^2\,\cos^2\zeta)} \, - \, \frac{1}{K} \, \left(1 \, - \, \frac{w_0}{1 \, - \, K^2\,\cos^2\zeta} \, \exp\left( -\frac{\tau_\infty}{\cos\zeta} \, - \, K\tau_\infty \right) \right) }{ \frac{2\,\gamma\,+\,K}{2\,\gamma\,-\,K} \, + \, \exp(-\,2\,K\,\tau_\infty) } \right)
 \end{align}

Finally one obtains the planetary albedo by putting the expressions for $a$ and $b$ above back into equation (12), evaluating equation (12) at the top of the atmosphere (${\rm \tau \, = \, \tau_\infty}$), and \textcolor{black}{dividing by} ${L\cos\zeta}$ \citep{TheBible}:

\begin{align}
    A_{pla} \, = \, \frac{2\,\gamma\,w_0}{(1\,-\,K^2\,\cos^2\zeta)} \, \left( \frac{2\,\gamma\,\cos\zeta \, + \, 1}{2\,\gamma \, - \, K} \, - \, \cos\zeta \right) \, + \, 2\,\gamma \, \left(1 \, - \, \frac{2\,\gamma\, + \, K}{2\,\gamma\, - \, K} \right) \, C. \label{eq:apla}
\end{align}

In the optically thin limit, expression (\ref{eq:apla}) tends to the cloud top reflectivity, imposed in equation (\ref{eq:BC_perfect_reflection}) to be $1$. In the optically thick limit, the expression reduces exactly back to the expression for the planetary albedo of a semi-infinite atmosphere \citep{TheBible}. In this case, the planetary albedo approaches 1 as ${w_0 \rightarrow 1}$, and approaches 0 as ${w_0 \rightarrow 0}$. We explore the full behavior of equation (\ref{eq:apla}) in the results section. 

\subsection{Numerical modelling of maximum planetary albedo \label{sec:methods_numerical}}

In order to retrieve the likely albedo range of K2-18b and further validate our analytic method, we employ numerical modelling that uses the retrieved properties of the above-cloud atmosphere. The albedo of K2-18b is calculated numerically by simulating radiative transfer of stellar (`shortwave') radiation in the above-cloud atmosphere. We use a line-by-line 1D radiative-transfer code, adapted from \citet{Wordsworth2021} to treat multiple-scattering, described fully in the appendix. The code calculates azimuthally averaged upwelling and downwelling spectral irradiances in a $\delta$-Eddington two-stream-type model \citep{Joseph1976}. We note that two-stream-type models are a broader class of radiative transfer models than the two-stream approximation itself, and the terminology refers to the method, not the number of streams considered. In this paper we effectively use 8-streams in a Gaussian quadrature procedure \citep{Wordsworth2017}. We calculate the reflectivity and transmissivity of atmospheric layers to `direct' (i.e., incident) and `diffuse' (i.e., scattered) radiation \citep{Briegleb1992}. Incoming shortwave radiation (ISR) provides the boundary condition on the incoming direct beam at the top of the atmosphere, and reflection from the cloud-top provides the boundary condition on the outgoing radiation at the base of the above-cloud atmosphere. The net outgoing shortwave radiation (OSR) is calculated from the outgoing radiation at the top of the atmosphere and the net planetary albedo is the ratio A$_{pla}$\,=\,OSR/ISR \citep{Wordsworth2017}. Further model details can be found in the appendix.

\begin{table}
\begin{center}
\caption{Input data for modelling the upper atmosphere of K2-18b \citep{MadhuCarbon}. Absorption data is taken from the {\sc Hitran2020} line list \citep{HITRAN2020}. Scattering data is taken from \citet{Cox} and \citet{SneepUbachs} (and references therein: \citet{Hohm,BideauMehu,Bates}). \label{tab:input_data_table}}
\begin{tabular}{l c c c}
\\\hline
Input data & \multicolumn{3}{c}{} \\\hline\hline
\emph{Species} & \multicolumn{2}{c}{\emph{Absorption}} & {\emph{Scattering}} \\
\ce{H2} & \multicolumn{2}{c}{{\sc{Hitran2020}} \citep{HITRAN2020}} & \citep{Cox}  \\
\ce{CH4} & \multicolumn{2}{c}{{\sc{Hitran2020}} \citep{HITRAN2020}} & \citep{SneepUbachs}  \\
\ce{CO2} & \multicolumn{2}{c}{{\sc{Hitran2020}} \citep{HITRAN2020}} & \citep{SneepUbachs}  \\
\ce{H2O} & \multicolumn{2}{c}{{\sc{Hitran2020}} \citep{HITRAN2020}} & \citep{Cox}  \\
\hline
\emph{Atmospheric parameters} & \multicolumn{3}{c}{\citep{MadhuCarbon}}\\
 & No Offset & 1 offset & 2 offsets \\
\ce{T} (K) & \,257$^{+127}_{-74}$ & \,242$^{+79}_{-57}$ & \,235$^{+78}_{-56}$  \\
log(P$_{\rm cloud}$ (bar)) & $-0.55^{+0.99}_{-1.20}$ & $-0.51^{+0.98}_{-1.20}$ & $-0.46^{+0.95}_{-1.15}$  \\
log(\ce{CH4}) & $-2.04^{+0.61}_{-0.72}$ & $-1.74^{+0.59}_{-0.69}$ & $-1.89^{+0.63}_{-0.70}$  \\
log(\ce{CO2}) & $-1.75^{+0.45}_{-1.03}$ & $-2.09^{+0.51}_{-0.94}$ & $-2.05^{+0.50}_{-0.84}$ \\
\textcolor{black}{log(\ce{H2O})} & \textcolor{black}{$<-3.21$} & \textcolor{black}{$<-3.06$} & \textcolor{black}{$<-3.49$} \\
\hline
\end{tabular}
\end{center}
\end{table}

We consider line absorption and Rayleigh scattering for \ce{H2}, \ce{CH4}, and \ce{CO2}, and \ce{H2}-\ce{H2} collisionally-induced absorption (Table \ref{tab:input_data_table}) to model the above-cloud atmosphere of K2-18b. We additionally consider the effect of \ce{H2O} in Figure \ref{fig:CH4_CO2_H2O} as a general result, but do not include \ce{H2O} in \textcolor{black}{our nominal models} of the above-cloud model of K2-18b since it will have condensed out deeper in the atmosphere than the pressure levels probed observationally. \textcolor{black}{Under the assumption that there is water in the deeper atmosphere, one would still expect there to be trace \ce{H2O} in the upper atmosphere according to the saturation vapour pressure, but at sufficiently low abundance to remain consistent with its non-detection \citep{MadhuCarbon}. Our estimates of maximum planetary albedo without any \ce{H2O} are therefore conservative upper limits when assessing the plausibility of a liquid water surface. We test the sensitivity of our results to the possible presence of trace \ce{H2O} at the detection limit from \citet{MadhuCarbon} in the Appendix.} A line-wing cutoff of 500\,cm$^{-1}$ is used for \ce{CO2} and 25\,cm$^{-1}$ for the remaining gases \citep{Wordsworth2017}. For the stellar SED input we use the spectrum of GJ436 (M3.5, T$_{\rm eff}$\,=\,3416\,K) from the MUSCLES Treasury survey \citep{France2016,Youngblood2016,Loyd2016}, as recommended by \citet{greg}. 

\section{Results \label{sec:results}}

\subsection{Analytic limits on planetary albedo}

\begin{figure}[htb!]
    \centering
    \includegraphics[width=0.8\columnwidth]{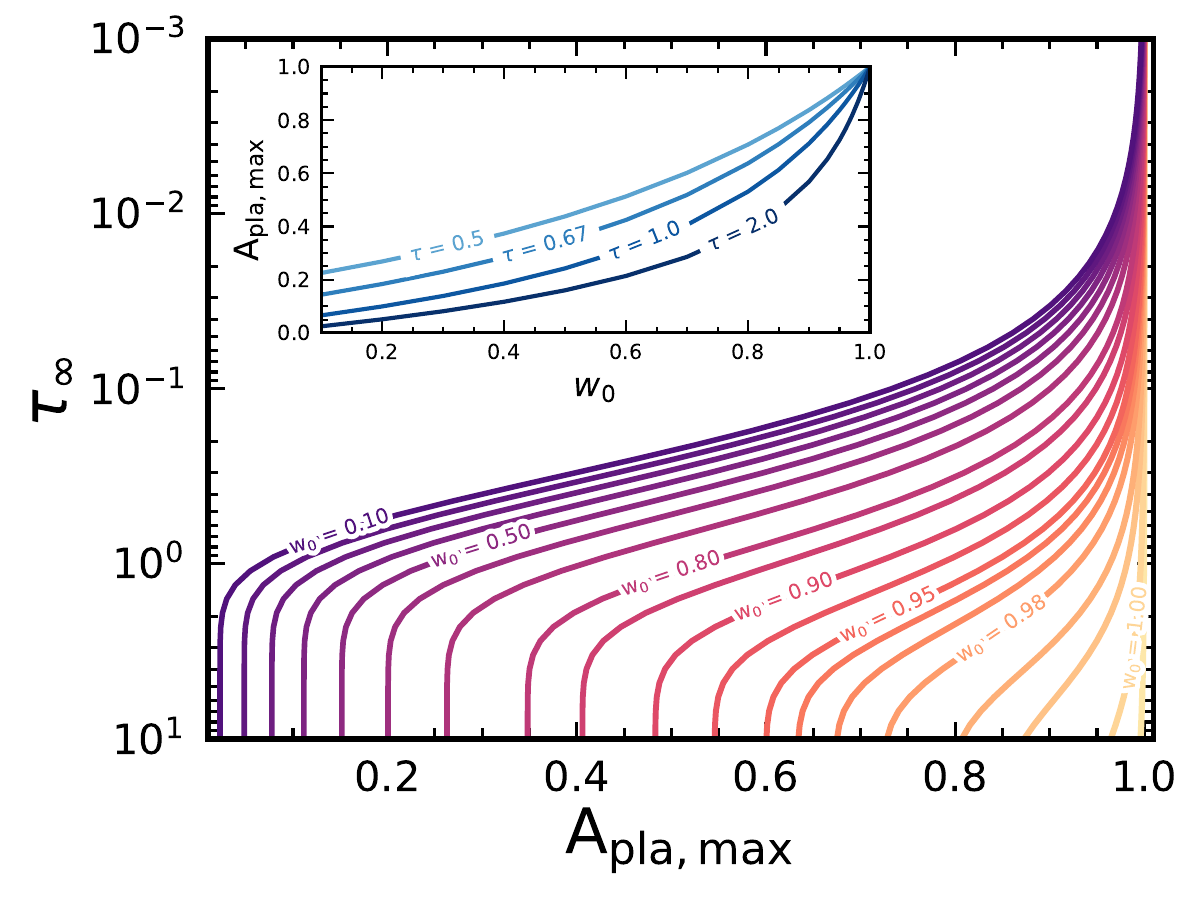}
    \caption{Analytic formulation of planetary albedo of a perfectly reflecting cloud layer resting at optical-depth $\tau_{\infty}$ from the top of the atmosphere, as a function of above-cloud scattering albedo $w_0$ (equation \ref{eq:apla}), with zenith angle $cos\zeta\,\sim\,\frac{2}{3}$. \textit{Inset:} maximum planetary albedo as a function of $w_0$ on contours of constant $\tau$.}
    \label{fig:analytic}
\end{figure}

In the limit of a planet possessing a perfectly reflective cloud layer, the resulting net planetary albedo still ranges between 0\,--\,1 as a function of the optical depth from the top of the atmosphere to the cloud-top, $\tau_{\infty}$, and the single-scattering albedo of the above-cloud atmosphere, $w_0$ (equation \ref{eq:apla}). The full behaviour of our analytic expression derived in the methods section is shown in Figure \ref{fig:analytic}. In the optically thin limit, the planetary albedo tends to the cloud albedo. In the optically thick limit, the planetary albedo ranges between 0\,--\,1 as a function of $w_0$. Since scattered radiation has further opportunity to be absorbed, only very high values of $w_0$ can lead to high planetary albedos in the limit of an optically thick above-cloud atmosphere.

Figure \ref{fig:analytic} can be used as a helpful first approximation to the maximum albedo of a planet given an observed transit spectrum. By definition, a transmission spectrum probes order unity optical depth in the atmosphere. However, since unity optical depth is probed in each wavelength bin respectively, a cloud top can rest at a range of possible values on the $\tau_{\infty}$ axis of Figure \ref{fig:analytic}. Cloud/haze truncation can cut off absorption features such that $\sim$\,unity optical depth in white light is due to the cloud/haze, not the gaseous opacity. The $\tau_{\infty}$ of the above-cloud atmosphere must then be less than unity and the maximum albedo is capable of being high. If clouds/hazes truncate the spectrum in between strong absorption features then the cloud top pressure is probed in wavelength windows where the atmosphere is otherwise relatively transparent to the stellar radiation, and so the cloud top must rest at some $\tau_{\infty}$ close to or greater than unity. The maximum albedo will then be high or low as a function of the above-cloud $w_0$, and one can read off the contours in Figure \ref{fig:analytic} to estimate the corresponding maximum albedo limit. 

\subsection{The observed atmosphere of K2-18b}

\begin{figure}[htb!]
    \centering
    \includegraphics[width=0.8\columnwidth]{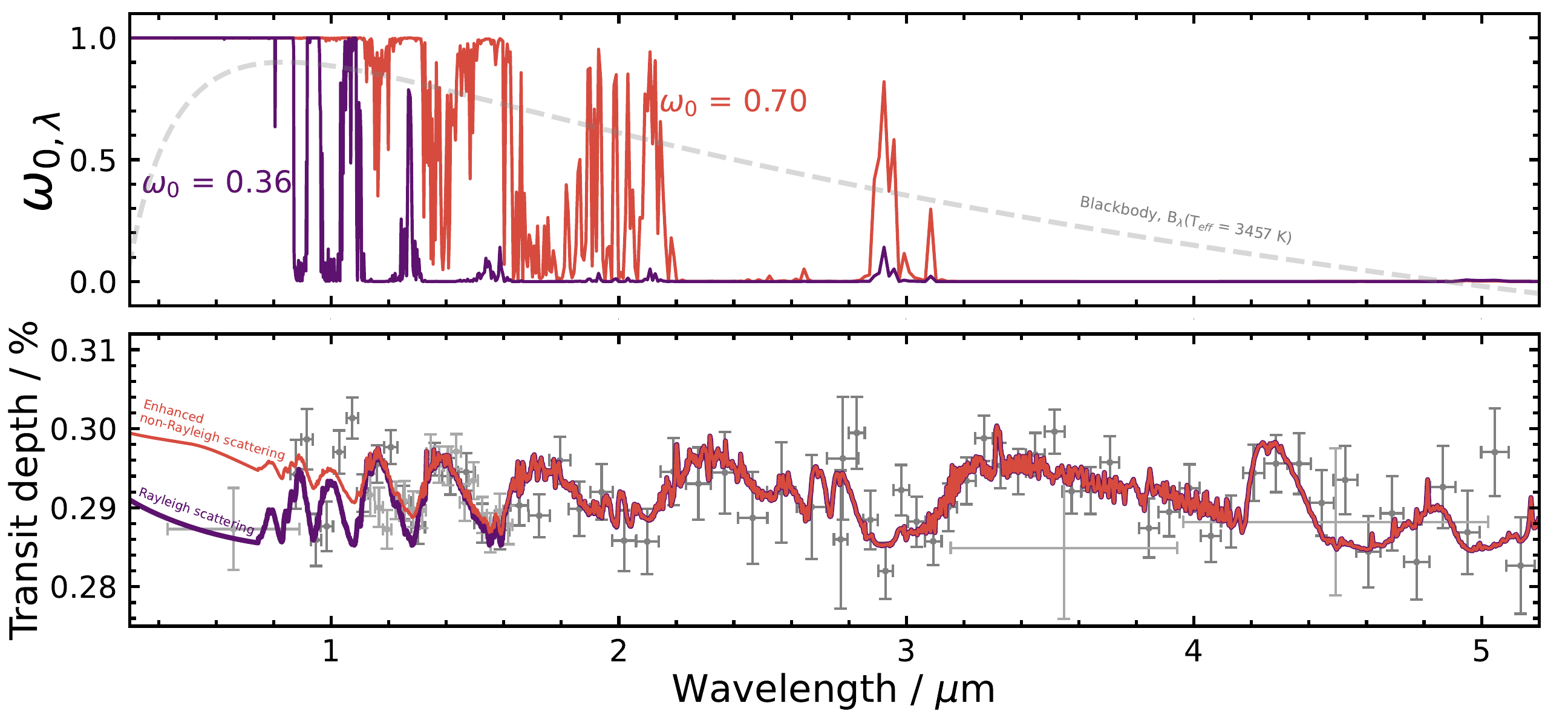}
    \caption{Wavelength dependent scattering albedo $w_0$ (\textit{top}), and transmission spectra (\textit{bottom}), for K2-18b with and without the enhanced non-Rayleigh haze parametrisation from \citet{MadhuCarbon}. Observational data points obtained with K2, HST, Spitzer, and JWST are shown in grey (\textit{bottom}) \citep{Benneke2019,MadhuCarbon}. }
    \label{fig:w0}
\end{figure}

In the case of K2-18b, the retrievals from \citet{reanalysis} found no evidence for clouds or hazes in the observed spectrum. The retrieval from \citet{MadhuCarbon} also found no evidence for any haze spectra, but reported weak model preference for a greatly enhanced non-Rayleigh-like haze parametrisation. The parametrisation assumes that hazes can be highly efficient scatterers, enhancing the scattering opacity by over 8 orders of magnitude in the retrieval \citep{MadhuCarbon}, and assumes that hazes can fit an arbitrary power law slope in the transmission spectrum at short wavelengths. The parametrisation is an \textit{ad hoc} approach to treating hazes and it is unclear what particles would achieve this degree of enhanced scattering \citep{Leconte2024}. However, stronger observational constraints on the scattering slope at short wavelengths would be necessary to definitively rule out this possibility \citep{reanalysis}. We therefore proceed with two models in estimating the albedo of K2-18b --- one where we consider only Rayleigh scattering from the above cloud atmosphere, and another with the inclusion of the enhanced non-Rayleigh slope parametrisation. \textcolor{black}{We use the same molecular abundance constraints for our calculations with and without the enhanced haze slope. The model preference for an enhanced haze slope in the retrieval of \citet{MadhuCarbon} is weak and constrained only at the shortest wavelengths, whereas the constraints for the molecular absorbers are relatively strong and constrained across the entire observing range for \ce{CH4}, and at longer wavelengths than those influenced by the enhanced haze slope for \ce{CO2}. It remains possible, however, that the molecular abundance constraints would differ slightly from the reported results of \citet{MadhuCarbon} if the enhanced haze slope was not included, which is a caveat to the calculations that we present without the enhanced haze parametrisation.} The resulting wavelength-dependent single-scattering albedos for each model are shown in Figure \ref{fig:w0}. The weighted average of the single scattering albedos, weighted by a unit black body curve at 3457\,K --- the effective temperature of K2-18 --- gives K2-18b's observable atmosphere a net $w_0$\,=\,0.36 for the nominal case, and $w_0$\,=\,0.70 for the enhanced non-Rayleigh parametrisation. 

Referring back to Figure \ref{fig:analytic} we can see that the resulting maximum possible planetary albedos that K2-18b could have for each possible $w_0$ depends on the $\tau_{\infty}$ parameter, but could only attain high enough albedos to deter the onset of Runaway Greenhouse (A$_{\rm pla}\,\gtrsim$\,0.6) if the above-cloud optical depth were below unity (Figure \ref{fig:analytic}, inset). To have an average $\tau_{\infty}$ below unity the spectrum would require a cloud deck that heavily truncates the observable absorption features. This cloud deck would raise the viewing level higher in altitude in the atmosphere where $\tau_\infty$ can be lower than order unity due to the molecular absorbers. This case is well exemplified by the observed transmission spectrum of GJ1214b, a notoriously very cloudy/hazy sub-Neptune planet whose transmission spectrum has consistently been observed to be flat \citep[e.g.,][]{Kempton2014}, and whose albedo has been calculated from phase curve analysis to be $\sim$\,0.51 \citep{Kempton2023} (notably less than that required by K2-18b to maintain liquid water beneath a 1\,bar \ce{H2} atmosphere).

Such cloud/haze truncation for K2-18b, at $\tau_{\infty}$ significantly below unity, appears to be inconsistent with the feature depths demonstrated in the observed spectrum (Figure \ref{fig:w0}). In contrast, a cloud deck resting at $\tau_{\infty}$ deeper than unity is consistent with the retrieved cloud top pressure being deeper than the retrieved reference pressure of the planetary radius \citep{MadhuCarbon}, and with the lack of a cloud deck at all in subsequent reanalysis \citep{reanalysis}. Taking the cloud deck to rest at $\tau_{\infty}$ deeper than unity, one can read off Figure \ref{fig:analytic} (inset) to find a maximum planetary albedo estimate. For the contours of $\tau_{\infty}\,\sim$\,0.67, we find A$_{\rm pla, max}\,<$\,0.26 for the nominal case, and A$_{\rm pla, max}\,<$\,0.52 for the enhanced non-Rayleigh parametrisation case. \textcolor{black}{The analytic treatment presented in Figure \ref{fig:analytic} is a useful toy model to demonstrate how it is possible to constrain planetary albedo based on only above-cloud atmospheric constraints, such as those obtained via transmission spectroscopy. Applying this approach in full to real systems, however, requires numerical modelling of the wavelength-dependent above-cloud optical depth. This depends on the inferred abundances of molecular absorbers and cloud top pressure level, and may vary significantly within the 3\,$\sigma$ observational error range.}
%The analytic treatment presented in Figure \ref{fig:analytic} thus provides a useful first approximation to the maximum possible albedo of K2-18b and a useful guide in general to constraining the albedo of any object in transmission spectroscopy. However the main uncertainty in applying this analytical approach to real systems remains in the mapping between retrieved composition and cloud top pressure level, to the total above-cloud optical depth, $\tau_{\infty}$. This mapping is non-trivial and will vary significantly within the 3\,$\sigma$ observational error range, and must be treated with numerical modelling.

\begin{figure}[b!]
    \centering
    \includegraphics[width=\columnwidth]{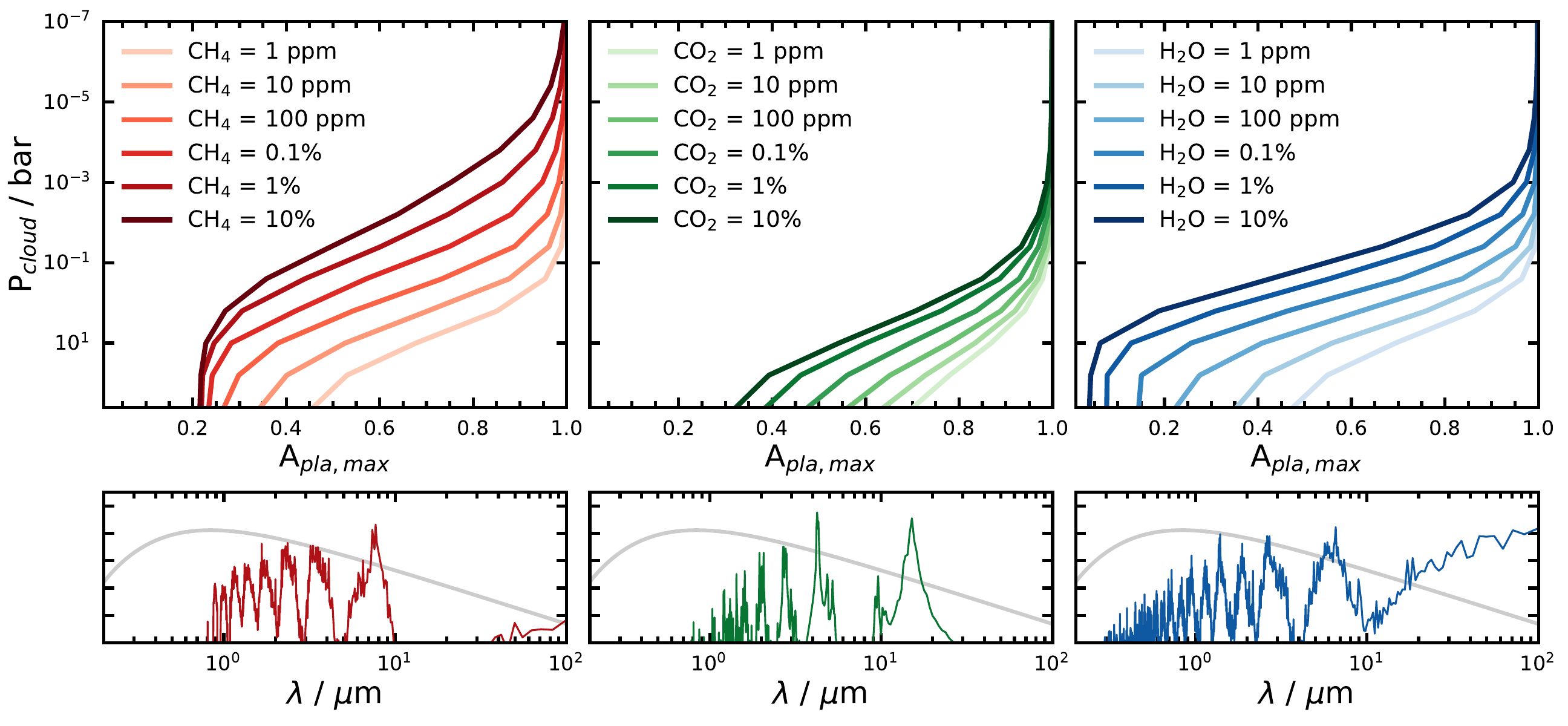}
    \caption{The planetary albedo for a \ce{H2} atmosphere containing \ce{CH4}, \ce{CO2}, or \ce{H2O} at abundances ranging from 1\,ppm to \textcolor{black}{10}\,\%, assuming a perfectly reflecting cloud top at pressure P$_{cloud}$\,(bar) (\textit{top}). The limit imposed on A$_{\rm pla}$ by above-cloud absorption depends on the absorption cross section of the molecular absorbers (\textit{bottom, coloured lines}) over the wavelength range of incident stellar radiation (\textit{bottom, grey lines}).}
    \label{fig:CH4_CO2_H2O}
\end{figure}

\subsection{Numerical limits on planetary albedo}

Using our numerical model, we demonstrate how above-cloud absorption in \ce{H2} atmospheres containing \ce{CH4}, \ce{CO2}, or \ce{H2O}, limits the maximum achievable planetary albedo, despite imposing perfect reflection from a cloud top (Figure \ref{fig:CH4_CO2_H2O}). \textcolor{black}{For increasing cloud-top pressure (i.e., deeper P$_{cloud}$ in Figure \ref{fig:CH4_CO2_H2O}), the maximum planetary albedo decreases in the same manner as the analytic formula (i.e., greater $\tau_{\infty}$ in Figure \ref{fig:analytic}), because there is more absorption opacity in the atmosphere above the clouds. Similarly,} the greater the abundance of a given molecular absorber in the atmosphere, the lesser the cloud-top pressure must be to attain the same maximum planetary albedo. At deep cloud-top pressures with high abundances of molecular absorbers, the effect saturates and the maximum planetary albedo asymptotes to a value set by the comparative strength of the atmospheric scattering compared to the molecular and continuum absorption. For a given atmospheric abundance of a species, the limit on albedo that the species imposes is strongest for \ce{H2O}, followed by \ce{CH4}, then \ce{CO2}. This is due to the relative strength of these species' respective absorption cross sections (Figure \ref{fig:CH4_CO2_H2O}, \textit{bottom}). %The limit imposed on planetary albedo by above-cloud \ce{CH4}, \ce{CO2}, and \ce{H2O} (Figure \ref{fig:CH4_CO2_H2O}) is well explained by our analytic description (Figure \ref{fig:analytic}). The mapping from retrieved cloud-top pressure to $\tau_{\infty}$ depends on the strength and abundance of the absorbing constituents of the atmosphere.  

The family of curves in our analytic model (Figure \ref{fig:analytic}) explain the behaviour in our numerical model (Figure \ref{fig:CH4_CO2_H2O}) qualitatively very well, and demonstrates the generality of the above-cloud albedo constraint. Importantly, the chemical constraints from observed exoplanet atmospheres requires no assumption to be made about the unseen deeper atmosphere in order to constrain planetary albedo. We now proceed with the specific physical and chemical constraints on the above-cloud atmosphere of K2-18b (Table \ref{tab:input_data_table}) to quantify its possible planetary albedo and compare to the required limits for potentially maintaining a liquid water ocean. \textcolor{black}{We assume a cloud coverage fraction of 1.0 so that our albedo calculations remain conservative upper limits. This also accounts for the possibility that the dayside-averaged cloud coverage fraction could be greater than the observed coverage fraction of $\sim$0.63\,--\,0.64 at the terminator region via transmission spectroscopy \citep{MadhuCarbon}}.

\begin{figure}[htbp]
    \centering
    \includegraphics[width=0.85\columnwidth]{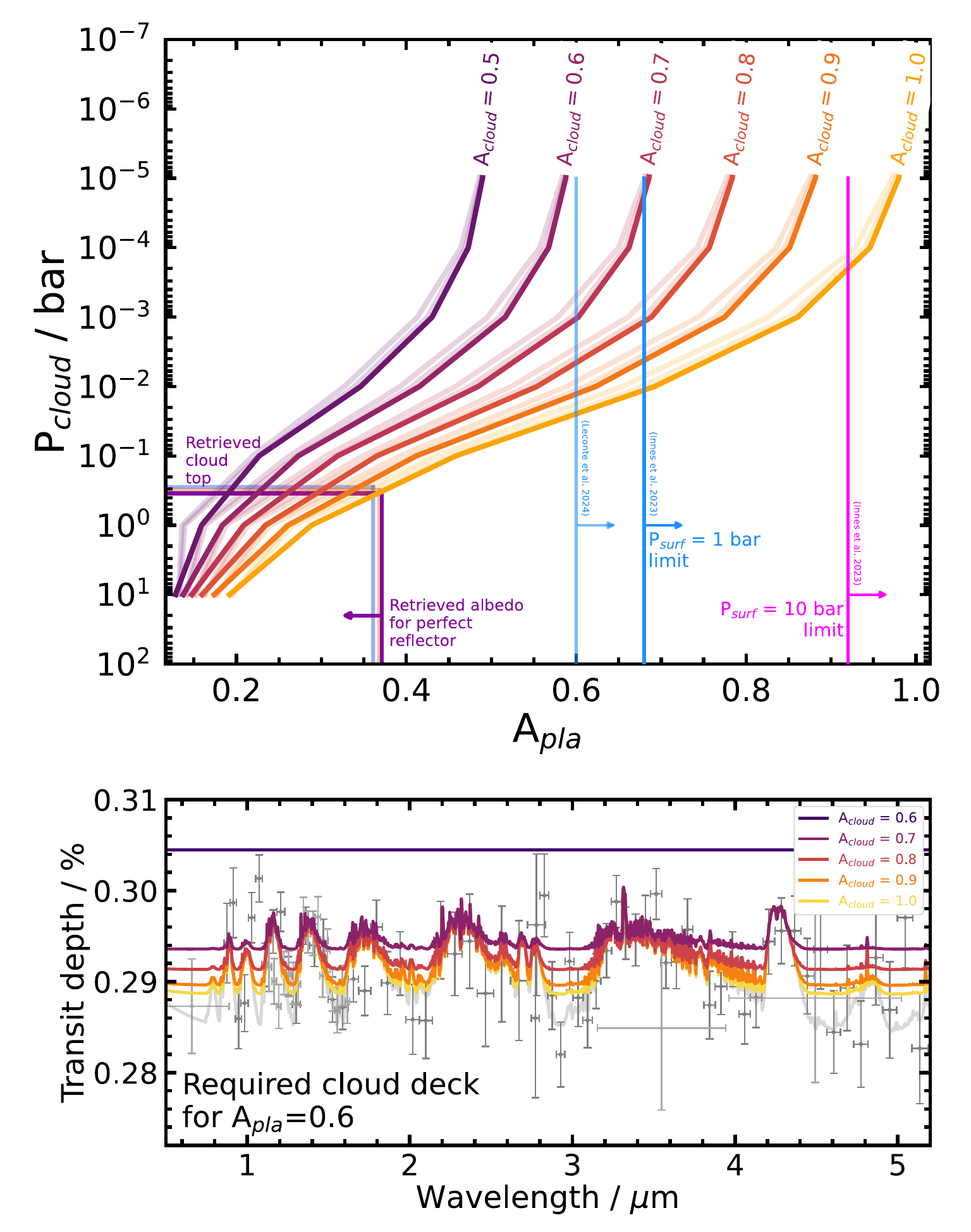}
    \caption{(\textit{Top}) Calculated albedo of K2-18b for the median values of P$_{cloud}$, T, \ce{CH4}, and \ce{CO2}, for three different instrumental offset cases (Table \ref{tab:input_data_table}; \citet{MadhuCarbon}). With a perfectly reflecting cloud layer, the retrieved parameters give A$_{pla}\,\lesssim$\,0.375, insufficient to keep a water surface stable against runaway greenhouse on K2-18b \citep{Leconte2024,Innes2023}. \textcolor{black}{(\textit{Bottom}) Spectral truncation that would occur for the required cloud top pressure to achieve A$_{pla}$\,=\,0.6, for full coverage by a grey cloud with albedo ranging from 0.6\,--\,1.0.}}
    \label{fig:K2-18b_is_dead}
\end{figure}

The median retrieval values of temperature, T\,(K), cloud-top pressure, P$_{\rm cloud}$\,(bar), \ce{CH4} mixing ratio, and \ce{CO2} mixing ratio, constrain K2-18b to have a maximum planetary albedo well below the required albedo to maintain a liquid water surface (Figure \ref{fig:K2-18b_is_dead}, \textit{top}). The planetary albedos calculated from the three different sets of possible instrumental offsets (Table \ref{tab:input_data_table}) all agree within $\sim$0.03 in planetary albedo. The maximum planetary albedo constraint, calculated with the assumption of a perfectly reflective cloud in the no-offset case (due to having the lowest \ce{CH4} estimate), is A$_{\rm pla}\,<\,$0.38. While the case of a perfect gray reflector (Figure \ref{fig:K2-18b_is_dead}, A$_{cloud}$\,=\,1.0) provides a useful upper limit, it is clearly unphysically high. Realistic values of the geometric albedos of cloud or haze layers that occur in nature are generally significantly lower than 1.0 due to both the intrinsic geometric reflectivity of the aerosol and their degree of sky coverage. For example, the `hydrogenated diamond' haze, which could be responsible for the high albedo of the cloudy planet GJ1214b, has an estimated geometric albedo of $\sim$\,0.6 \citep{Ohno2024}. \textcolor{black}{For a similar cloud/haze layer to be capable of maintaining an ocean surface on K2-18b under only 1\,bar of \ce{H2}, the required cloud top pressure would lead to a flat transit spectrum (Figure \ref{fig:K2-18b_is_dead}, \textit{bottom}, A$_{cloud}$\,=\,0.6).}

However physical and chemical properties retrieved from planetary atmospheres have associated 1\,$\sigma$ error ranges which will map through to Bayesian confidence intervals on the posterior parameter estimation. We can therefore proceed by marginalising over parameter uncertainties on T, P$_{\rm cloud}$, \ce{CH4}, and \ce{CO2}, and over input cloud albedos between 0\,--\,1, to `retrieve' a posterior on the planetary albedo itself. Retrieving the albedo of a planet in this way is possible in principle for any planet that has constraints on its observable atmosphere, but has not previously been formalised. 

Marginalising over the parameter uncertainties provides not a maximum albedo constraint but instead a posterior distribution on the \textit{likely} planetary albedo given the data. 
Obtaining a posterior on the planetary albedo in this way is akin to retrieving any other atmospheric parameter. It does, however, differ from other retrieval constraints in transmission spectroscopy because it requires calculating the ratio of reflected incident stellar radiation from a vertical column of atmosphere, averaging over incidence angle with multiple streams in Gaussian quadrature. Whereas, retrievals from transmission spectroscopy calculate atmospheric opacities in a chord through the planet's atmosphere.

We marginalise over the retrieval posteriors by drawing parameter values from their reported probability distributions \citep{MadhuCarbon}. We sample retrieval values of temperature, T\,(K), cloud-top pressure, P$_{\rm cloud}$\,(bar), \ce{CH4} mixing ratio, and \ce{CO2} mixing ratio, from Gaussian distributions according to the retrieved constraints (Table \ref{tab:input_data_table}; \citet{MadhuCarbon}). We sample the unknown possible cloud albedo from a uniform prior distribution ranging 0\,--\,1. We treat the parameter uncertainties as independently distributed \textcolor{black}{whereas in a retrieval, parameters will often be correlated with each other, for example cloud depth and abundance of molecular absorbers. The implication of independently sampling from the reported parameter distributions is that our posterior distributions for planetary albedo will be wider and reach farther extremes than in the case of correlated uncertainties. In absence of available retrieval data, we proceed with the assumption of independent parameter uncertainties.}

\begin{figure}[htbp]
    \centering
    \includegraphics[width=0.8\columnwidth]{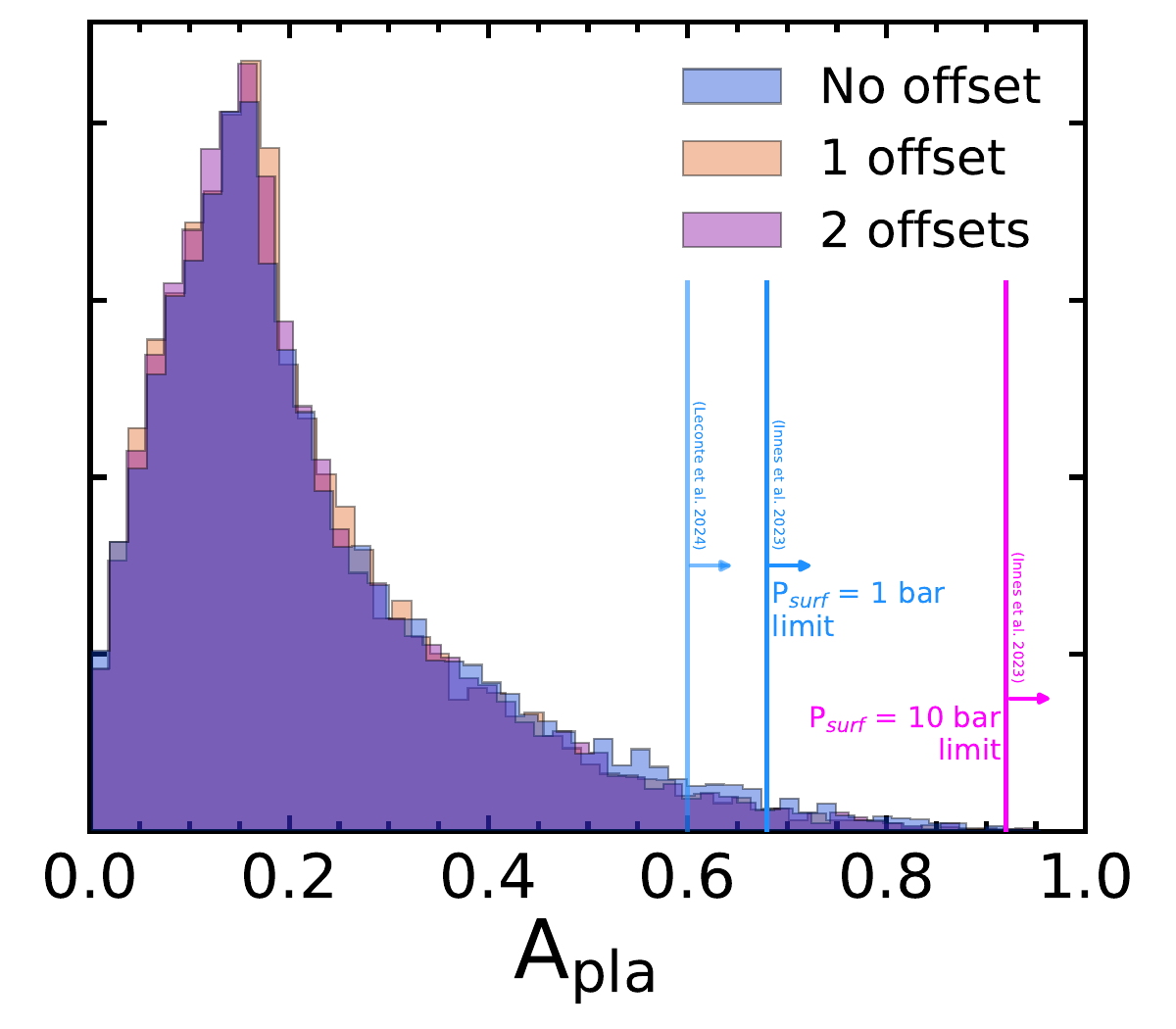}
    \caption{Posterior distribution of planetary albedo A$_{\rm pla}$ marginalising over possible cloud albedo (A$_c\,\epsilon$\,[0,\,1]) and the reported error ranges for P$_{\rm cloud}$, T, \ce{CH4}, and \ce{CO2}, for three different instrumental offset cases (Table \ref{tab:input_data_table}; \citet{MadhuCarbon}). The median planetary albedo A$_{\rm pla}\sim$\,0.17\,--\,0.18.}
    \label{fig:hist_Apla}
\end{figure}

The resulting median planetary albedos thus obtained are A$_{\rm pla}\sim$\,0.17\,--\,0.18 for the three offset cases (Figure \ref{fig:hist_Apla}, Appendix). \textcolor{black}{When trace \ce{H2O} at the detection limit is also included the median values drop to A$_{\rm pla}\sim$\,0.13\,--\,0.14 (Figure \ref{fig:hist_Apla_water}).} The requirements for planetary albedo inferred from modelling studies of the runaway greenhouse limit in \ce{H2}-dominated atmospheres are $\gtrsim$\,0.60 \citep{Leconte2024} and $\gtrsim$\,0.68 \citep{Innes2023} for 1\,bar of \ce{H2}, and $\gtrsim$\,0.92 \citep{Innes2023} for 10\,bars of \ce{H2}. \textcolor{black}{Based on the measured mass and radius of K2-18b, interior structure modelling from \citet{Madhu2020} found a minimum surface pressure of $\sim$\,4\,bar for planetary structures consistent with an ocean layer, with surface pressures lower than 4\,bar only possible when pushing out to the 1\,$\sigma$ error thresholds. This suggests that the minimum required albedo is likely to be even greater than the reported 1\,bar limits from \citet{Innes2023} and \citet{Leconte2024}.}% \citet{Madhu2020} note that the minimum allowable surface pressure for the ocean case may be lower than 4\,bar for model solutions that remain admissible within the 1\,$\sigma$ uncertainties on K2-18b's measured mass and radius.}

\begin{figure}[hb!]
    \centering
    \includegraphics[width=0.42\columnwidth]{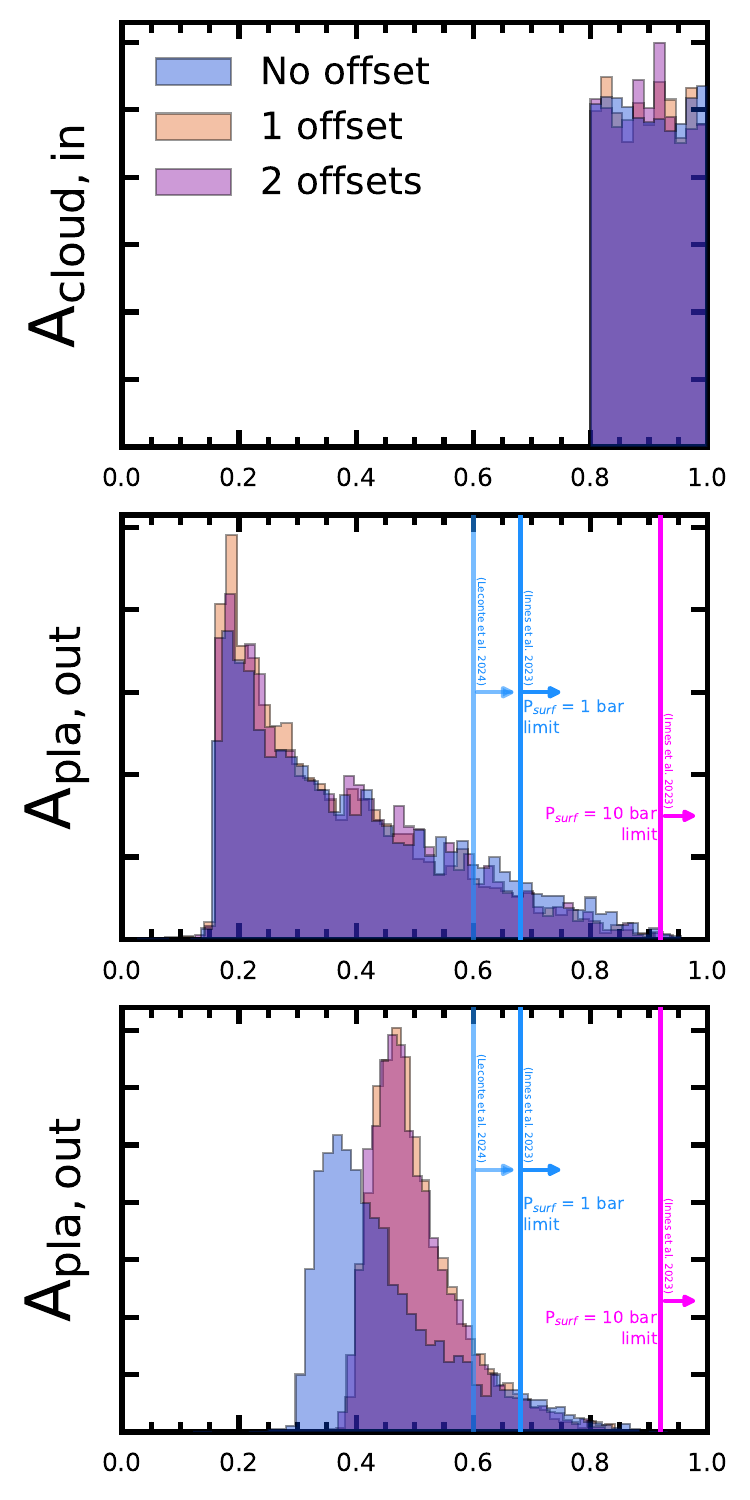}
    \caption{Posterior distribution of planetary albedo A$_{\rm pla}$ marginalising over the reported error ranges for P$_{\rm cloud}$, T, \ce{CH4}, and \ce{CO2}, for very high cloud albedos (A$_c\,\epsilon$\,[0.8,\,1.0], \textit{top panel}). The median planetary albedos range A$_{\rm pla}\sim$\,0.31\,--\,0.35 for the nominal case (\textit{middle panel}). When the enhanced non-Rayleigh-like haze parametrisations are included (Figure \ref{fig:w0}), the median planetary albedos range A$_{\rm pla}\sim$\,0.42\,--\,0.49 (\textit{bottom panel}). The parametrisation scales the Rayleigh scattering opacity by factors of $10^{7.31}$, $10^{8.20}$, and $10^{8.21}$, and has wavelength dependence $\lambda^{-11.67}$, $\lambda^{-11.11}$, and $\lambda^{-11.34}$, for no offset, 1 offset, and 2 offsets respectively \citep{MadhuCarbon}. However, we note that previous studies have found no haze spectra from the atmosphere of K2-18b, and studies have cautioned against such high scaling factors for the scattering opacity \citep{Leconte2024}. Subsequent reanalysis also found no evidence for the haze parametrisations \citep{reanalysis}.} %so we present their corresponding A$_{\rm pla}$ values with caution here as conservative over-estimates. }
    \label{fig:hist_Apla_high_Ac}
\end{figure}

Reconciling the observational data with the interpretation of an ocean surface underlying only 1 bar of \ce{H2} would require not only a tightly constrained and high value of the cloud albedo prior, but also simultaneously pushing out to the 3\,$\sigma$ limits on the retrieved atmospheric abundance ratios of absorbing species and cloud-top pressure (Figure \ref{fig:hist_Apla_high_Ac}). Isolating input albedos of A$_{\rm cloud}\,\epsilon$\,[0.8,\,1.0] (Figure \ref{fig:hist_Apla_high_Ac}, \textit{top}), we see that the highest input cloud albedos map onto planetary albedo posteriors with medians A$_{\rm pla}\,\sim$\,0.31\,--\,0.35 (Figure \ref{fig:hist_Apla_high_Ac}, \textit{middle}). Including also the greatly enhanced haze parametrisation (Figure \ref{fig:w0}) brings these median planetary albedos up to 0.42\,--\,0.49 (Figure \ref{fig:hist_Apla_high_Ac}, \textit{bottom}), still below currently published limits required for preventing runaway greenhouse under the thinnest \ce{H2} atmospheres \citep{Innes2023,Leconte2024}. These numerical A$_{\rm pla, max}$ values are all consistent with the upper limit derived from our analytic expression for the enhanced haze case, A$_{\rm pla, max}\,<$\,0.52 (Figure \ref{fig:analytic}). 

\begin{table}
\begin{center}
\caption{Median planetary albedo outputs from Figures \ref{fig:hist_Apla}, \ref{fig:hist_Apla_high_Ac}, \ref{fig:hist_Apla_water} and \ref{fig:hist_Apla_high_Ac_water}. \label{tab:output_all}}
\begin{tabular}{l c c c}
\\\hline\hline
Case & \multicolumn{3}{c}{Albedo}\\\hline
 &  &  & (A$_{cloud}\,\epsilon\,$[0.8,\,1] plus \\
 & (A$_{cloud}\,\epsilon\,$[0,\,1]) & (A$_{cloud}\,\epsilon\,$[0.8,\,1]) & enhanced scattering) \\
 & A$_{\rm pla}$ & A$_{\rm pla}$ & A$_{\rm pla}$ \\\hline\hline
No Offset & 0.176 & 0.347 & 0.420 \\
1 offset & 0.172 & 0.312 & 0.485 \\
2 offsets & 0.170 & 0.330 & 0.483 \\\hline
\textcolor{black}{including trace \ce{H2O} (see Appendix)} &  &  &  \\\hline\hline
\textcolor{black}{No Offset} & \textcolor{black}{0.140} & \textcolor{black}{0.301} & \textcolor{black}{0.402} \\
\textcolor{black}{1 offset} & \textcolor{black}{0.129} & \textcolor{black}{0.277} & \textcolor{black}{0.478} \\
\textcolor{black}{2 offsets} & \textcolor{black}{0.133} & \textcolor{black}{0.270} & \textcolor{black}{0.482} \\
\hline\hline
\end{tabular}
\end{center}
\end{table}

\section{Discussion \label{sec:discussion}}

Our results demonstrate how planetary albedo is fundamentally constrained by the above-cloud atmosphere in a way that can be conveniently calculated from spectroscopic observation of exoplanets, and used to accurately quantify their potential habitability. The qualitative and quantitative agreement between our analytic derivation and full numerical treatment demonstrates that our conclusions are not model specific, and can be widely applied to any target with constraints on the physical and chemical properties of their observable atmospheres. Constraining the maximum planetary albedo from the above-cloud atmosphere also does not require any assumption to be made about the deeper atmosphere, relying only upon the region of the atmosphere that is probed in remote observation. 

\subsection{Broader application}

Observations of the above-cloud atmosphere most strongly constrain planetary albedo for deep cloud tops (lying at high pressure), where there is a high abundance of molecular absorbers, and for planets around cool stars. Sub-Neptune exoplanets that can be observed currently with JWST have high atmospheric scale heights, percent level or even greater concentrations of common greenhouse gases, and orbit M-dwarfs. Thus, while our method can place albedo constraints on any target with sufficient constraints on their upper atmospheres, the method will be optimally applicable to sub-Neptunes orbiting M-dwarfs, which JWST is observationally well poised for. 

An albedo constraint can be calculated from either transmission or emission spectroscopy with JWST, each with different benefits and limits. In transmission spectroscopy, shorter wavelengths can be probed allowing $w_0$ to be constrained, and the cloud top pressure can be estimated from truncation of the spectrum, allowing $\tau_{\infty}$ to be constrained. Since transmission spectroscopy probes the terminator region of an exoplanet's atmosphere, the albedo constraint requires assuming that the day-side atmosphere is well represented by the terminator region. This is appropriate for planets with efficient heat redistribution by atmospheric circulation, acting to homogenise clouds and hazes. The constraint would differ if a planet were to have inefficient circulation and cloud accumulation exclusively at the substellar point that dispersed before the terminator. Conversely, the inefficient heat redistribution would lead to higher required albedos for maintaining a given day-side surface temperature. In emission spectroscopy, constraints can be made on the molecular absorbers and cloud/haze emission continua, enabling potentially better constraints on $\tau_{\infty}$, but weaker constraints on $w_0$, since the longer wavelengths are generally less sensitive to haze scattering properties. Planetary albedo estimates from emission spectroscopy benefit, however, from providing measurements from the full day-side rather than only the terminator region.

\subsection{K2-18b and habitability}

\textcolor{black}{Applying our method to K2-18b as a case study demonstrates that the median retrieval constraints reported in \citet{MadhuCarbon} require that K2-18b is highly unlikely to be habitable, even if it only possessed a \ce{H2} atmosphere as thin as 1\ bar with a perfectly reflecting global cloud layer. In our planetary albedo posteriors with cloud albedo ranging between 0\,--\,1, only $\sim$3\% of the distribution without enhanced scattering, and $\sim$3.5\% of the distribution with enhanced scattering, lies above A$_{\rm pla}\gtrsim$\,0.60 \citep{Leconte2024}. In these extreme models, high planetary albedos are produced by having, on average, input cloud albedos of $\sim$0.9 and cloud top pressures $\sim$10$^{-2}$\,bar. In the case of a 10\,bar atmosphere, vanishingly small proportions of the albedo posterior attain the requisite A$_{\rm pla}\gtrsim$\,0.92 \citep{Innes2023}: only $\sim$0.02\% without the enhanced scattering, and $\sim$0.004\% with the enhanced scattering. The proportion of enhanced scattering models with A$_{\rm pla}\gtrsim$\,0.92 is lower than without the enhanced scattering because multiple scattering provides more opportunity for radiation to be absorbed, which reduces albedos that could otherwise have been extremely high due to clouds. These percentages have also been calculated assuming independent parameter posteriors. If parameter correlations were considered then the planetary albedo distributions would be narrower and the percentages we quote would be even lower. We conclude that the observed data disfavour the possibility of a surface water ocean on K2-18b.} %\,---\,lower than the nominal case because the distribution with enhanced scattering is more strongly peaked around the median
%In our albedo calculations without enhanced haze scattering, $\sim$3\% of the posterior distribution lies above the A$_{\rm pla}\gtrsim$\,0.60 threshold \citep{Leconte2024} and $\sim$0.02\% lies above the A$_{\rm pla}\gtrsim$\,0.92 threshold \citep{Innes2023}. With enhanced haze scattering, $\sim$3.5\% of the posterior distribution lies above the A$_{\rm pla}\gtrsim$\,0.60 threshold \citep{Leconte2024} and $\sim$0.004\% lies above the A$_{\rm pla}\gtrsim$\,0.92 threshold \citep{Innes2023} due to being more strongly peaked around the median. The observed data therefore disfavour the possibility of a surface water ocean on K2-18b.} 
%This agrees with other constraints from the water\,:\,rock fraction that planet formation theory can attain \citep[e.g.,][]{Lodders2003}, and the theory of atmospheric escape from planetary atmospheres \citep[e.g.,][]{Luger2015}, each of which independently require fine-tuning to achieve the requirements for a habitable ocean on K2-18b \citep{Wogan2024}. Additionally, the albedo values that we obtain (Table \ref{tab:output_all}) are overestimates compared to the albedo values that would be obtained from the reanalysis of the JWST observations \citep{reanalysis}. In the reanalysis, no evidence was found for clouds or hazes, and the retrieved methane abundance was greater, each of which will significantly reduce K2-18b's maximum attainable planetary albedo (Figure \ref{fig:CH4_CO2_H2O}). We thus conclude, via only the observations themselves, that K2-18b is not a liquid water habitable world.

%%%

Reconciling the observed transmission spectrum of K2-18b with the interpretation of being an ocean world \textcolor{black}{requires significant fine tuning of a cloud/haze layer, which would also have to be constituted of a material with near-perfectly reflective properties across NIR wavelengths but exhibit no} greenhouse warming of the atmosphere beneath the clouds. \textcolor{black}{Whether any substance exists in nature that can satisfy these constraints around an M-dwarf without causing IR scattering-greenhouse warming of the planet \citep{TheBible} is questionable.} The limiting scenario outlined above also only applies to the case of K2-18b possessing a 1\,bar atmosphere. If K2-18b has a $>$\,1\,bar atmosphere\textcolor{black}{, as suggested by the interior structure models of \citet{Madhu2020},} then even this limiting scenario \textcolor{black}{may be} incapable of giving K2-18b a sufficient albedo to maintain a liquid water ocean. 

Previous work has called into question the assumption that K2-18b has a 1\,bar atmosphere of \ce{H2}. The first problem that arises is from planet formation: in order to possess an ocean layer but only 1\,bar or less of \ce{H2}, K2-18b requires a water\,:\,rock ratio which exceeds all current limits from planet formation theory \textcolor{black}{\citep[e.g.,][]{Lodders2003,Marcus2010,LuquePalle2022,Nixon2024}}. For example, with a water\,:\,rock ratio of $\sim$\,90\,\%, K2-18b would need a \ce{H2} envelope of $\sim$\,130\,bar \citep{Madhu2020}. In order to possess an atmosphere of \ce{H2} as thin as $\sim$\,1\,bar, K2-18b therefore requires a water fraction $>$\,90\,\% of the total planet mass. The upper limit to the water\,:\,rock ratio that can be achieved in planet formation beyond the ice line, however, is \textcolor{black}{expected to be $\lesssim$\,50\,\% \citep[e.g.,][]{Lodders2003,LuquePalle2022,Nixon2024}, although limits as high as $\lesssim$\,75\,\% have been considered previously \citep{Marcus2010}.} The second problem arises from the early evolution of the K2-18 system, in which K2-18b will have sustained 10s\,--\,100s Myrs of super-luminous irradiation from its host star K2-18 \citep{Luger2015}. For K2-18b's initial massive \ce{H2} envelope to have been eroded away to 1\,bar or less, yet not eroded beyond the limit imposed by the observed transmission spectrum itself, has been described previously as `fortuitous' \citep{Wogan2024}. The third problem is the continued long-term evolution of the K2-18 system, whereby K2-18b would have had to retain 1\,bar of \ce{H2} (and no more) over geological timescales despite the high extreme UV flux of M-dwarfs compared to G-dwarfs such as the Sun \citep{Wogan2024}. Hence, for K2-18b to somehow maintain a substantial atmosphere that is no more than 1\,bar of \ce{H2} overlying a water ocean for geological timescales, there must be an unknown regulatory feedback for \ce{H2} atmospheres in contact with ocean reservoirs that drives the system to possess 1\,bar of \ce{H2}.

\textcolor{black}{Since K2-18b receives a lower instellation flux than other currently-characteriseable sub-Neptunes \citep{Madhu2021}, our results suggest that the rest of the observed sub-Neptune population are also unlikely to be able to maintain liquid water oceans. The requirement for very high cloud/haze albedo to stabilise habitable water ocean conditions under \ce{H2}-dominated atmospheres creates a fundamental challenge to characterisation of these hypothetical worlds: evidence of such habitable conditions in transmission spectroscopy will be obscured from view by the very clouds and hazes that ensure the habitable conditions in the first place. Current observing facilities would therefore struggle to provide characterisation of a genuinely habitable sub-Neptune among the current short-period planets discovered.}

Our results also demonstrate how mere consistency with water condensation is not sufficient evidence for an observed exoplanet to be claimed as (conventionally) habitable: K2-18b has an equilibrium temperature consistent with the condensation of water somewhere in its atmosphere and yet cannot maintain a surface ocean. The surge of interest in a potentially habitable K2-18b inspired by the initial JWST observations \citep{MadhuCarbon} points to the challenge that the community faces in identifying conditions accepted as showing signs of habitability and life. Such claims will be sure to occur more frequently now with cumulative observations from JWST, and will be critical to address for future missions that are optimised for the search for habitable and potentially inhabited exoplanets, such as the Habitable Worlds Observatory \citep[HWO;][]{HWO} and Large Interferometer for Exoplanets \citep[LIFE;][]{Sascha}.

\subsection{Could K2-18b possess a magma ocean?}

Since K2-18b likely receives too much incident stellar flux to maintain a liquid water ocean layer beneath a \ce{H2}-dominated atmosphere, alternative interior structures with hotter surfaces or deep atmospheres are favoured by the observational data. Two physically plausible possibilities for the interior structure of K2-18b are the magma ocean scenario and the mini-Neptune, or `gas dwarf', scenario. In the magma ocean scenario, atmospheres with mass fractions between \textcolor{black}{$\sim$\,10$^{-2}$\,--\,6.2\,\%} of the total planet mass are consistent with the bulk density for a range of plausible core-mass fractions \textcolor{black}{and only trace water content} \citep{Madhu2020}. For the surface gravity and bulk parameters of K2-18b, these limits correspond to surface pressures between \textcolor{black}{$\sim$\,10$^{4}$\,--\,10$^{5}$\,bar. Lower atmospheric mass fractions and surface pressures are also feasible in scenarios with significant mass fractions of \ce{H2O} dissolved in the melt.} At the surface, the gas-phase chemistry would be conditioned by the magma-ocean\,--\,atmosphere interaction. In the limit of a deeper atmosphere, the interior structure approaches that of the mini-Neptune scenario, where there is no discrete density change across an atmosphere-mantle interface, but instead matter is extensively mixed as a supercritical envelope, like the deep interiors of ice giant or gas giant planets \citep{HelledFortney}. 

\citet{ShorttleMagma} reconciled the apparent discrepancy between the fact that K2-18b was not in its `habitable zone' and the observed lack of \ce{NH3} in its transmission spectrum by demonstrating that a chemically reducing magma ocean would naturally sequester atmospheric N and lead to low \ce{NH3} abundances in the atmosphere. \ce{NH3} non-detection is thus not sufficient in general to evidence an ocean surface. \citet{ShorttleMagma} show how a wide region of the possible parameter space for the magma-ocean case would be capable of explaining the observed transmission spectrum within the observational error on the abundances of atmospheric carbon species. In particular, they highlight how the corresponding \ce{CO}/\ce{CO2} ratio in the atmosphere above a chemically reduced magma ocean was always greater than 1 and could potentially be a tracer for magma ocean chemistry. Even in this case, the stronger absorption features of \ce{CO2} still mean that \ce{CO2} would dominate over CO with respect to transmission feature height. Alternatively, a magma ocean would also fit the observed spectrum with respect to $\ce{CH4}$ detection and $\ce{NH3}$ non-detection without any $\ce{CO2}$ or CO features \citep{ShorttleMagma}, which was the atmospheric composition found in the spectral reanalysis by \citet{reanalysis}.

The results of \citet{ShorttleMagma} were also reproduced later by \citet{FrancesMagma} who modelled K2-18b as a magma ocean world. The results presented in the main article did not provide a good fit to the transmission spectrum because the analysis focused on one input gas composition for the H-C-N-S system, of 50$\times$\,Solar metallicity. When, in their Appendix B they explored models for 100$\times$ and 300$\times$\,Solar metallicity, they found similar results to those highlighted above: that a magma ocean on K2-18b could lead to low atmospheric \ce{NH3} abundances and predicts a ratio of CO/\ce{CO2} greater than 1 \citep{ShorttleMagma}. \citet{FrancesMagma} did not find models that fit both the depletion of \ce{NH3} and the presence of \ce{CO2} simultaneously because the analysis only considered scalings of the Solar metallically for the H-C-N-S system. However, non-Solar elemental ratios are a plausible outcome of planet formation (e.g., as on the Earth), \textcolor{black}{and an expected outcome of magma-ocean\,--\,atmosphere interactions \citep[e.g.,][]{Werlen2025}}, providing a wide potential compositional space sub-Neptunes may occupy.

Like the magma-ocean scenario, a mini-Neptune or `gas dwarf' scenario would also provide a plausible fit to the observational data \citep{Wogan2024}. The terminology `mini-Neptune' refers to idea that the planet resembles the interior structure of the Solar System ice giants insofar as there is a thick envelope where the gaseous atmosphere continuously increases in pressure and density into and beyond a supercritical phase of matter deeper in the interior. In this scenario there would still certainly be a rocky/metallic core of a significant planetary mass fraction. The difference between the magma-ocean and the mini-Neptune scenarios relates to whether there is a distinction between the mantle and envelope, be that clearly defined or `fuzzy' \citep{Kite2019}, versus a more Neptune-like interior structure \citep{HelledFortney}. The magma ocean scenario and mini-Neptune scenario are thus not necessarily dissimilar, but exist on a continuum. We encourage deeper exploration of magma-ocean chemistry both experimentally and with numerical modelling in order to constrain what place on the spectrum from magma-ocean to mini-Neptune K2-18b and other presently characteriseable sub-Neptunes, such as TOI-270d \citep{Benneke2024,Mans} and GJ9827d \citep{GJ9827d}, occupy. The era of JWST may provide a wealth of data on magma-ocean\,--\,atmosphere interaction, particularly on sub-Neptune planets. The requirement for very low instellations for sub-Neptunes to potentially harbour liquid water oceans means that the discovery of such planets will require future missions such as the HWO \citep{HWO} or LIFE \citep{Sascha,Leung2025}.

\section{Conclusions \label{sec:conclusions}}

Planetary albedo is fundamentally important to assessing the potential habitability of exoplanets, but is not a free parameter when transmission or emission spectroscopic data exist. We here demonstrate how planetary albedo is limited by the above-cloud atmosphere, which is observationally accessible via transit spectroscopy observations. We derive an analytic expression to demonstrate how the planetary albedo of a planet with a perfectly reflecting cloud layer nonetheless depends on the optical depth and scattering versus absorbing properties of the above-cloud atmosphere. We apply our model to the much debated case of K2-18b, proposed to be a potentially habitable planet with a liquid water ocean underlying a \ce{H2} dominated atmosphere. We use a multiple-scattering, line-by-line numerical radiative transfer model to constrain the planetary albedo of K2-18b. The retrieved albedo is found to lie at values below the threshold required for K2-18b to host a habitable liquid water ocean, based solely on the observational data. Marginalising over parameter uncertainties gives median planetary albedos of 0.17\,--\,0.18. For only very high input cloud albedos ([0.8\,--\,1.0]), the median albedo for K2-18b becomes 0.31\,--\,0.35. Including, in addition, a greatly enhanced haze parametrisation, the medians become 0.42\,--\,0.49. All calculated planetary albedos fall below the threshold required to maintain a water ocean on K2-18b under even only 1\,bar of \ce{H2}. K2-18b is therefore not a potentially habitable planet, and sub-Neptunes currently being observed with JWST are likely to be magma ocean or `gas dwarf' worlds. The methods demonstrated in this paper are equally applicable to any planet with constraints on the physical and chemical properties of the above-cloud atmosphere, enabling the quantification of their potential habitability with current observing capabilities.

\section*{Acknowledgements}

Thanks to Caroline Dorn for helpful discussions about planet formation. S.J. acknowledges funding support from ETH Zurich and the NOMIS Foundation in the form of a research fellowship. The NOMIS Foundation ETH Fellowship Programme and respective research are made possible thanks to the support of the NOMIS Foundation. O.S. acknowledges support from UKRI(STFC) grant UKRI1184.

\bibliography{refs}{}
\bibliographystyle{aasjournal}

\appendix

\section{Including trace water}

\textcolor{black}{Here we test the sensitivity of our results to the inclusion of trace \ce{H2O} in the above-cloud atmosphere of K2-18b. We repeat figures \ref{fig:hist_Apla} and \ref{fig:hist_Apla_high_Ac} with \ce{H2O} included at a mixing ratio consistent with its 2$\sigma$ non-detection limit (Table \ref{tab:input_data_table}; \citet{MadhuCarbon}).}

\begin{figure}[htbp]
    \centering
    \includegraphics[width=0.8\columnwidth]{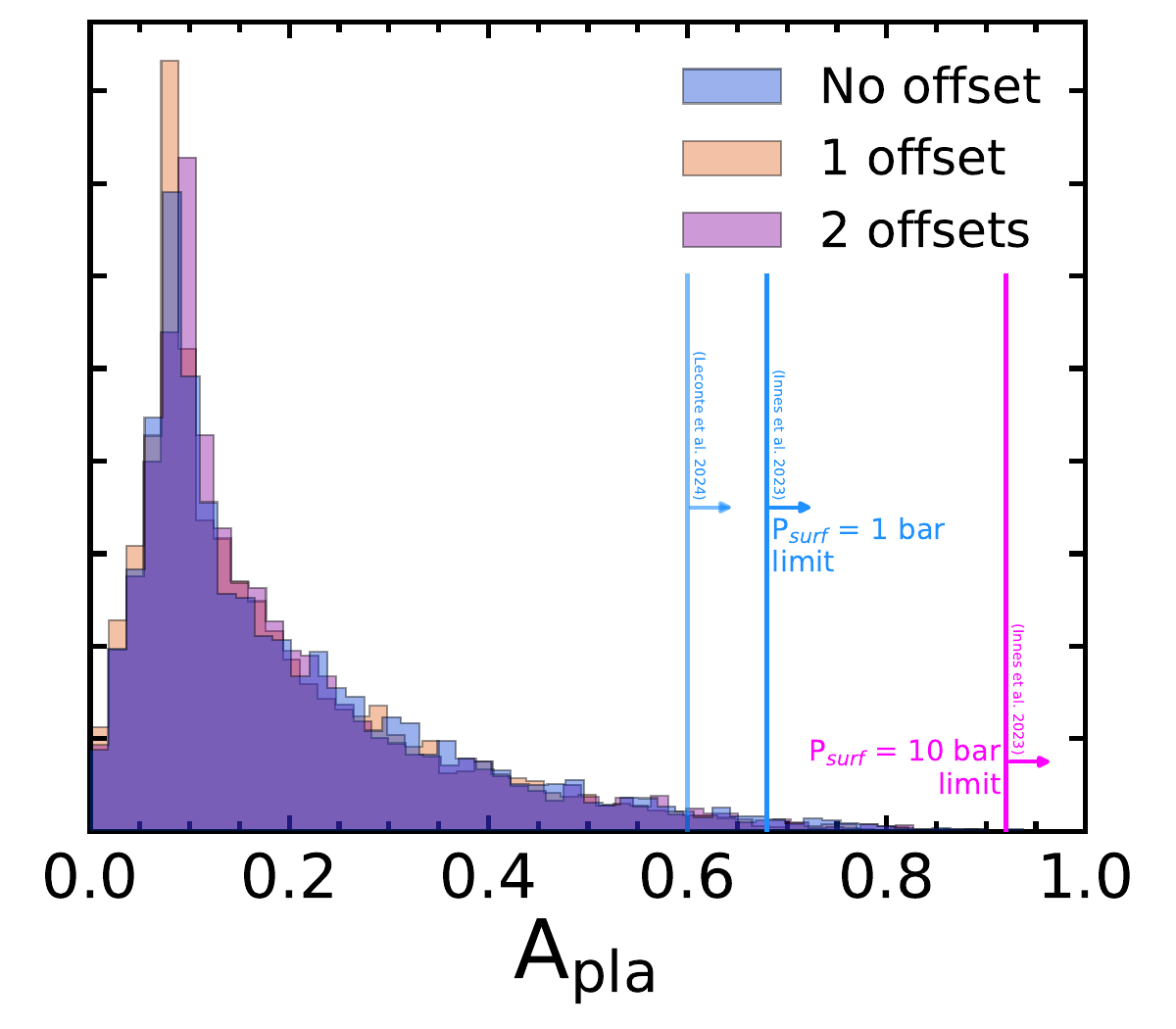}
    \caption{Posterior distribution of planetary albedo A$_{\rm pla}$ marginalising over possible cloud albedo (A$_c\,\epsilon$\,[0,\,1]) and the reported error ranges for P$_{\rm cloud}$, T, \ce{CH4}, and \ce{CO2}, for three different instrumental offset cases, including \ce{H2O} at the 2$\sigma$ non-detection limit (Table \ref{tab:input_data_table}; \citet{MadhuCarbon}). The median planetary albedo A$_{\rm pla}\sim$\,0.13\,--\,0.14.}
    \label{fig:hist_Apla_water}
\end{figure}

\begin{figure}[htbp]
    \centering
    \includegraphics[width=0.42\columnwidth]{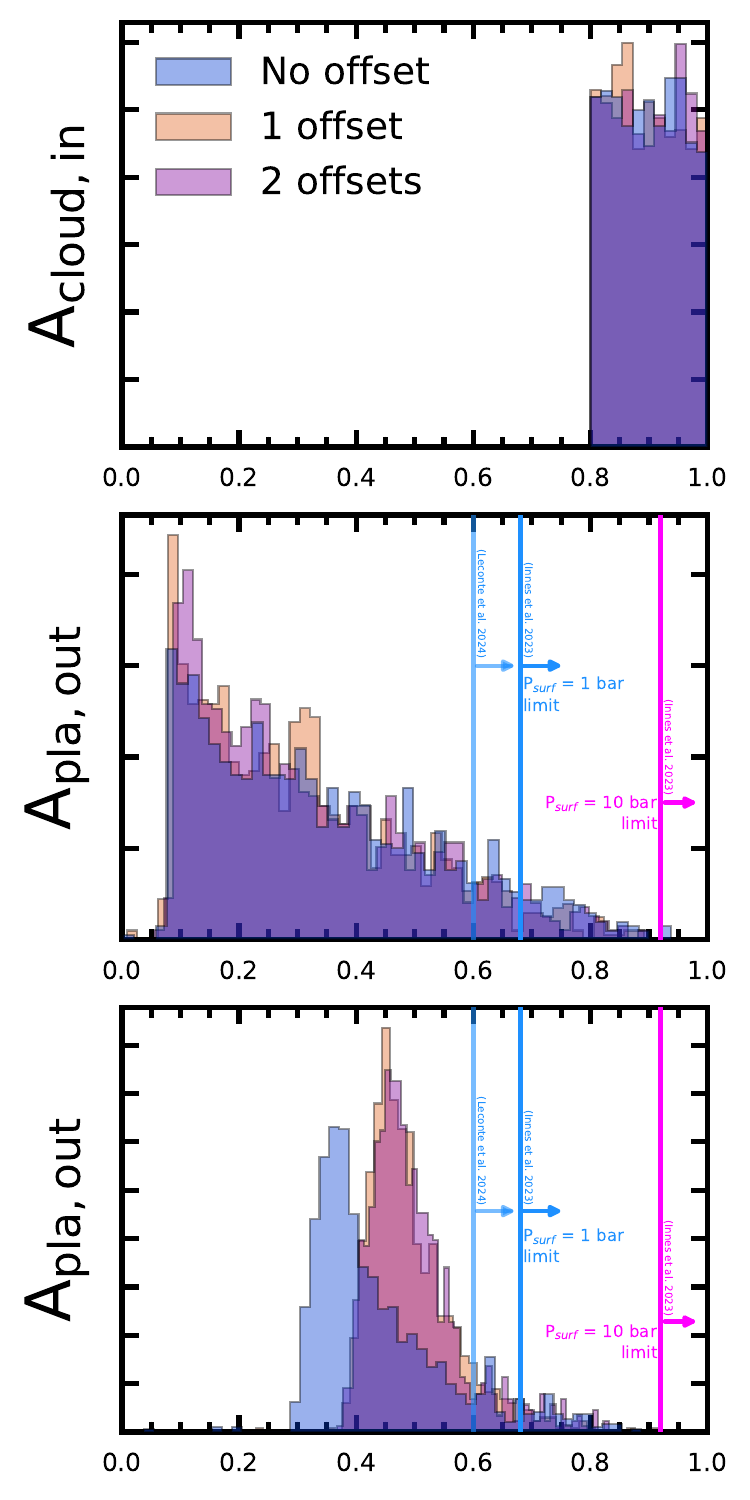}
    \caption{Posterior distribution of planetary albedo A$_{\rm pla}$ marginalising over the reported error ranges for P$_{\rm cloud}$, T, \ce{CH4}, and \ce{CO2}, for very high cloud albedos (A$_c\,\epsilon$\,[0.8,\,1.0], \textit{top panel}), including \ce{H2O} at the 2$\sigma$ non-detection limit (Table \ref{tab:input_data_table}; \citet{MadhuCarbon}). The median planetary albedos range A$_{\rm pla}\sim$\,0.27\,--\,0.30 for the nominal case (\textit{middle panel}). When the enhanced non-Rayleigh-like haze parametrisations are included (Figure \ref{fig:w0}), the median planetary albedos range A$_{\rm pla}\sim$\,0.40\,--\,0.48 (\textit{bottom panel}). The parametrisation scales the Rayleigh scattering opacity by factors of $10^{7.31}$, $10^{8.20}$, and $10^{8.21}$, and has wavelength dependence $\lambda^{-11.67}$, $\lambda^{-11.11}$, and $\lambda^{-11.34}$, for no offset, 1 offset, and 2 offsets respectively \citep{MadhuCarbon}. However, we note that previous studies have found no haze spectra from the atmosphere of K2-18b, and studies have cautioned against such high scaling factors for the scattering opacity \citep{Leconte2024}. Subsequent reanalysis also found no evidence for the haze parametrisations \citep{reanalysis}.}
    \label{fig:hist_Apla_high_Ac_water}
\end{figure}

\newpage

\section{Numerical radiative transfer model}

Vertical optical depth, $\tau$, at a given wavenumber, $\nu$, is defined as:
\begin{equation}
	\tau = \frac{\kappa(p_{s}-p)}{g}
\end{equation}
for pressure $p$, surface gravity $g$, and mass absorption coefficient $\kappa$ \citep{Wordsworth2017}. These equations are discretised and solved over evenly spaced layers in log-pressure.

The azimuthal averaging is discretised using 8-point Gaussian quadrature, averaged using the Gaussian weighting procedure:
\begin{equation}
	F_{up} = 2\pi \int_{0}^{1}I\mu d\mu \approx 2\pi \sum_{i} I(\tau_{i})\mu_{i} w_i
    \label{eq:gauss}
\end{equation}
for Gaussian weights $w_i$ \citep{Wordsworth2017}.

Since we are interested here in how planetary albedo is influenced by the above-cloud atmosphere, we need only model the stellar flux (`shortwave') in the above-cloud atmosphere, with reflection from the cloud top imposed at the base of the above-cloud atmosphere. This cloud top rests at the retrieved cloud-top pressure level. The temperature in the above-cloud atmosphere is assumed to be isothermal at the retrieved temperature, and only influences the results via the temperature dependence of absorption features. 

We model radiative transfer of shortwave radiation following \citet{Briegleb1992}. This radiative transfer procedure has also been more recently applied in \citet{Mendonca2015} and \citet{Jordan2025}.  The parametrisation first calculates the optical depth $\tau$, single scattering albedo $\omega$, asymmetry parameter $g$, and forward scattering fraction $f$, within each atmospheric layer, accounting for absorbing and scattering constituents (denoted subscript $i$), according to the equations: 
\begin{equation}
    \tau = \sum_{i}\tau_{i}
\end{equation}
\begin{equation}
    \omega = \frac{\sum_{i}\omega_{i}\tau_{i}}{\tau}
\end{equation}
\begin{equation}
    g = \frac{\sum_{i}g_{i}\omega_{i}\tau_{i}}{\omega\tau}
\end{equation}
\begin{equation}
    f = \frac{\sum_{i}f_{i}\omega_{i}\tau_{i}}{\omega\tau}
\end{equation}
These parameters are rescaled with the $\delta$ adjustment, according to the $\delta$-Eddington approximation \citep{Joseph1976}, by removing the fraction of scattered energy associated with the forward-scattered peak:
\begin{equation}
    \tau^{*} = \tau(1-\omega f)
\end{equation}
\begin{equation}
    \omega^{*} = \omega\frac{1-f}{1-\omega f}
\end{equation}
\begin{equation}
    g^{*} = \frac{g-f}{1-f}
\end{equation}
We detail the model in its full generality here, however we note that the results we present have $g=0$ and $f=0$ since the atmospheric scattering contributions come only from Rayleigh Scattering or an enhancement of Rayleigh-like scattering.

With these rescaled parameters, the fraction of spectral radiance transmitted through a layer, the \textit{transmissivity}, and the fraction reflected by a layer, the \textit{reflectivity}, can be computed for every atmospheric layer. Incoming radiation from the host star is initially direct, with angle of incidence $\mu_{o}$. Once scattered, radiation is assumed to be diffuse and isotropic, as opposed to transmitted direct radiation which continues to propagate as direct radiation with angle of incidence $\mu_{o}$. The radiation field in the atmosphere therefore has two components to it: \textit{direct} radiation, with an angular dependence, and \textit{diffuse} radiation, which is isotropic. For every layer, one thus needs to describe the transmissivity and reflectivity to \textit{direct} radiation ($T$ and $R$), and the transmissivity and reflectivity to \textit{diffuse} radiation ($\overline{T}$ and $\overline{R}$) as described in \citet{Briegleb1992}:
\begin{equation}
    R(\mu_{o}) = (\alpha - \gamma)\overline{T} e^{-\frac{\tau^{*}}{\mu_{o}}} + (\alpha + \gamma)\overline{R} - (\alpha - \gamma)
\end{equation}
\begin{equation}
    T(\mu_{o}) = (\alpha - \gamma)\overline{R} e^{-\frac{\tau^{*}}{\mu_{o}}} + (\alpha + \gamma)\overline{T} - (\alpha + \gamma -1) e^{-\frac{\tau^{*}}{\mu_{o}}}
\end{equation}
\begin{equation}
    \overline{R}(\mu_{o}) = (u+1)(u-1)(e^{\lambda \tau^{*}} - e^{-\lambda\tau^{*}})N^{-1}
\end{equation}
\begin{equation}
    \overline{T}(\mu_{o}) = 4uN^{-1}
\end{equation}
for $\alpha$, $\gamma$, $\lambda$, N, and u defined as:
\begin{equation}
    \alpha = \frac{3}{4}\omega^{*}\mu_{o} \frac{1+g^{*}(1-\omega^{*})}{1-\lambda^{2}\mu_{o}^{2}}
\end{equation}
\begin{equation}
    \gamma = \frac{1}{2}\omega^{*} \frac{1+3g^{*}(1-\omega^{*})\mu_{o}^{2}}{1-\lambda^{2}\mu_{o}^{2}}
\end{equation}
\begin{equation}
    N = (u+1)^{2}e^{\lambda \tau^{*}} - (u-1)^{2}e^{-\lambda \tau^{*}}
\end{equation}
\begin{equation}
    u = \frac{3(1-\omega^{*}g^{*})}{2\lambda}
\end{equation}
\begin{equation}
    \lambda = \sqrt{3(1-\omega^{*})(1-\omega^{*}g^{*})}
\end{equation}
The reflectivity and transmissivity to direct and diffuse radiation in a given layer can then be used to calculate the reflectivity and transmissivity to direct and diffuse radiation at every layer interface using the `adding-layer' method \citep{Briegleb1992}. The adding-layer method makes two passes through the atmosphere on every iteration: on the downwards pass, moving from the top of the atmosphere downwards, reflectivity and transmissivity of layers are subsequently combined with the reflectivity and transmissivity of the entire column above until the base of the atmosphere is reached; on the upwards pass, layers are combined subsequently with the column below until the top of the atmosphere is reached. The combination of two overlying layers (or a layer combined to the column below or above), with layer 1 overlying layer 2, is calculated as:
\begin{equation}
    R_{12}(\mu_{o}) = R_{1}(\mu_{o})+\frac{\overline{T}_{1}((T_{1}(\mu_{o})-e^{-\frac{\tau_{1}^{*}}{\mu_{o}}})\overline{R}_{2} + e^{-\frac{\tau_{1}^{*}}{\mu_{o}}}R_{2}(\mu_{o}))}{1-\overline{R}_{1}\overline{R}_{2}}
\end{equation}
\begin{equation}
	T_{12}(\mu_{o}) = e^{-\frac{\tau_{1}^{*}}{\mu_{o}}}T_{2}(\mu_{o}) + \frac{\overline{T}_{2}((T_{1}(\mu_{o})-e^{-\frac{\tau_{1}^{*}}{\mu_{o}}}) + e^{-\frac{\tau_{1}^{*}}{\mu_{o}}}R_{2}(\mu_{o})\overline{R}_{1})}{1-\overline{R}_{1}\overline{R}_{2}}
\end{equation}
\begin{equation}
	\overline{R}_{12} = \overline{R}_{1} + \frac{\overline{T}_{1}\overline{R}_{2}\overline{T}_{1}}{1-\overline{R}_{1}\overline{R}_{2}}
\end{equation}
\begin{equation}
	\overline{T}_{12} = \frac{\overline{T}_{1}\overline{T}_{2}}{1-\overline{R}_{1}\overline{R}_{2}}
\end{equation}
The formulas above provide the combined reflectivities and transmissivities for upwelling and downwelling radiation at every layer interface. At a layer interface, with rescaled optical depth $\tau^{*}$ from the top of the atmosphere to the interface: $e^{-\frac{\tau^{*}}{\mu_{o}}}$ describes the direct beam transmission from the top of the atmosphere the the interface; $R_{up}(\mu_{o})$ describes the reflectivity of the entire column below the interface to direct radiation incident from above; $T_{dn}(\mu_{o})$ describes the total transmission of the entire column above the interface to radiation incident from above; $\overline{R}_{up}$ describes the reflectivity of the entire column below the interface to diffuse radiation from above; $\overline{R}_{dn}$ describes the reflectivity of the entire column above the interface to diffuse radiation from below. The upwelling and downwelling spectral fluxes can then be evaluated at every layer interface according to:
\begin{equation}
	F_{up} = \frac{e^{-\frac{\tau^{*}}{\mu_{o}}}R_{up}(\mu_{o}) + (T_{dn}(\mu_{o})-e^{-\frac{\tau^{*}}{\mu_{o}}})\overline{R}_{up}}{1-\overline{R}_{dn}\overline{R}_{up}}
\end{equation}
\begin{equation}
	F_{dn} = e^{-\frac{\tau^{*}}{\mu_{o}}} + \frac{(T_{dn}(\mu_{o})-e^{-\frac{\tau^{*}}{\mu_{o}}}) + e^{-\frac{\tau^{*}}{\mu_{o}}}R_{up}(\mu_{o})\overline{R}_{dn}}{1-\overline{R}_{dn}\overline{R}_{up}}
\end{equation}
The resulting upward and downward spectral fluxes are summed to obtain spectrally integrated fluxes. Accounting for the angular discretiseation in Gaussian quadrature (equation \ref{eq:gauss}), the final planetary albedo is then the net ratio of the outgoing shortwave flux to the incident shortwave flux at the top of the atmosphere.

\end{document}